\documentclass[12pt,preprint]{aastex}
\usepackage{epsfig}
\usepackage{natbib}
\usepackage{graphicx}
\usepackage{mathrsfs,amssymb}
\usepackage{amsmath}
\newcommand	\beq	{\begin{equation}}	
\newcommand	\eeq	{\end{equation}}	
\newcommand       \Angstrom     {\,{\rm \AA}}

\newcommand       \cm           {\,{\rm cm}}
\newcommand       \mm           {\,{\rm mm}}

\newcommand       \g            {\,{\rm g}}

\newcommand       \Gyr      {\,{\rm Gyr}}

\newcommand       \nH           {n_{\rm H}}
\newcommand       \NH           {N_{\rm H}}
\newcommand       \NHI          {N({\rm HI})}
\newcommand       \NHH         {N({\rm H_2})}
\newcommand       \simlt        {\lesssim}
\newcommand       \simgt        {\gtrsim}

\newcommand       \um           {\mu{\rm m}}
\newcommand       \mum          {\,{\rm \mu m}}
\newcommand       \ppm          {\,{\rm ppm}}

\newcommand       \mH           {m_{\rm H}}

\newcommand       \simali       {\sim\,}
\newcommand       \magni        {\,{\rm mag}}
\newcommand       \rmH          {{\rm H}}
\newcommand       \HH           {\,{\rm H}}
\newcommand       \Htwo        {{\rm H}_2}
\newcommand       \rhosil        {\rho_{\rm sil}}
\newcommand       \rhogra        {\rho_{\rm gra}}

\newcommand	  \cism         {\left[{\rm C/H}\right]_{\rm ISM}}

\newcommand	  \oism         {\left[{\rm O/H}\right]_{\rm ISM}}
\newcommand	  \feism        {\left[{\rm Fe/H}\right]_{\rm ISM}}
\newcommand	  \mgism        {\left[{\rm Mg/H}\right]_{\rm ISM}}
\newcommand	  \siism        {\left[{\rm Si/H}\right]_{\rm ISM}}

\newcommand	  \cdust        {\left[{\rm C/H}\right]_{\rm dust}}

\newcommand	  \odust        {\left[{\rm O/H}\right]_{\rm dust}}

\newcommand	  \cgas        {\left[{\rm C/H}\right]_{\rm gas}}

\newcommand	  \sgas        {\left[{\rm S/H}\right]_{\rm gas}}
\newcommand	  \ogas        {\left[{\rm O/H}\right]_{\rm gas}}
\newcommand	  \fegas       {\left[{\rm Fe/H}\right]_{\rm gas}}
\newcommand	  \mggas       {\left[{\rm Mg/H}\right]_{\rm gas}}
\newcommand	  \sigas       {\left[{\rm Si/H}\right]_{\rm gas}}
\newcommand	  \otot        {\left[{\rm O/H}\right]_{\rm tot}}

\newcommand	  \muc         {\mu_{\rm C}}
\newcommand	  \muo         {\mu_{\rm O}}
\newcommand	  \musi        {\mu_{\rm Si}}
\newcommand	  \mufe        {\mu_{\rm Fe}}
\newcommand	  \mumg        {\mu_{\rm Mg}}

\newcommand	  \Vdust      {V_{\rm dust}}
\newcommand	  \Vsil         {V_{\rm sil}}
\newcommand	  \Vgra        {V_{\rm gra}}

\newcommand	  \VFe          {V_{\rm Fe}}
\newcommand	  \VFeOa     {V_{\rm FeO}}
\newcommand	  \VFeOb     {V_{\rm Fe_2O_3}}
\newcommand	  \VFeOc     {V_{\rm Fe_3O_4}}

\def    \rhosil		{\rho_{\rm sil}}
\def    \rhogra		{\rho_{\rm gra}}
\def    \rhoFe		{\rho_{\rm Fe}}
\def    \rhoFeOa	{\rho_{\rm FeO}}
\def    \rhoFeOb	{\rho_{\rm Fe_2O_3}}
\def    \rhoFeOc	{\rho_{\rm Fe_3O_4}}

\def    \dof		{{\rm dof}}
\def    \xo             {x_{\rm o}}
\def      \AV          {{A_V}}
\def      \RV          {{R_V}}
\def      \AUU        {{A_U}}
\def      \AB          {{A_B}}
\def      \AJ          {{A_J}}
\def      \AH         {{A_H}}
\def      \AK          {{A_K}}
\def    \AintKK	    {A_{\rm int}^{\rm\scriptsize KK}}
\def    \Aintobs   {A_{\rm int}^{\rm obs}}
\def    \AintobsWD   {A_{\rm int}^{\rm obs}(\rm\scriptsize WD)}%
\def    \AintobsWLJ   {A_{\rm int}^{\rm obs}(\rm\scriptsize WLJ)}
\DeclareMathAlphabet{\mathsc}{OT1}{cmr}{m}{sc}
\def\testbx{bx}%
\DeclareRobustCommand{\ion}[2]{%
\relax\ifmmode
\ifx\testbx\f@series
{\mathbf{#1\,\mathsc{#2}}}\else
{\mathrm{#1\,\mathsc{#2}}}\fi
\else\textup{#1\,{\mdseries\textsc{#2}}}%
\fi}
\countdef\decade=200
\advance\decade by \year
\countdef\hours=201
\advance\hours by \time
\divide\hours by 60
\countdef\mins=202
\advance\mins by \hours
\multiply\mins by 60
\multiply\hours by 100
\countdef\miltime=203
\advance\miltime by \hours
\advance\miltime by \time
\advance\miltime by -\mins
\def\today{\number\decade.\number\month.\number\day.\number\miltime}
\shorttitle{Interstellar Extinction and Elemental Abundances}
\title{Interstellar Extinction and Elemental Abundances
\\{\small DRAFT: \today ~~}}

\author{Wenbo Zuo\altaffilmark{1,2,3}, 
        Aigen Li\altaffilmark{3} 
        and Gang Zhao\altaffilmark{1,2}}
\altaffiltext{1}{CAS Key Laboratory of Optical Astronomy, 
       National Astronomical Observatories, 
       Chinese Academy of Sciences, 
       Beijing 100101, China;
       {\sf gzhao@nao.cas.cn}}
\altaffiltext{2}{School of Astronomy and Space Science, 
       University of Chinese Academy of Sciences, 
       Beijing 100049, China}
\altaffiltext{3}{Department of Physics and Astronomy,
                  University of Missouri,
                  Columbia, MO 65211, USA;
                  {\sf lia@missouri.edu}}

\begin{document}
\begin{abstract}
Elements in the interstellar medium (ISM) 
exist in the form of gas or dust.
The interstellar extinction and elemental abundances 
provide crucial constraints on
the composition, size and quantity of interstellar dust.
Most of the extinction modeling efforts 
have assumed the total (gas and dust) abundances
of the dust-forming elements---known as 
the ``interstellar abundances'',  
``interstellar reference abundances'',
or ``cosmic abundances''---to be solar 
and the gas-phase abundances
to be environmentally independent.
However, it remains unclear if the solar abundances
are an appropriate representation 
of the interstellar abundances.
Meanwhile, the gas-phase abundances 
are known to exhibit appreciable variations 
with local environments. 
Here we explore the viability of
the abundances of B stars, 
the solar and protosolar abundances,
and the protosolar abundances augmented 
by Galactic chemical enrichment (GCE)
as an appropriate representation 
of the interstellar abundances
by quantitatively examining the extinction 
and abundances of ten interstellar sightlines
for which both the extinction curves
and the gas-phase abundances of all the major 
dust-forming elements (i.e., C, O, Mg, Si and Fe) 
have been observationally determined. 
Instead of assuming a specific dust model
and then fitting the observed extinction curves, 
for each sightline we apply the model-independent 
Kramers-Kronig relation, 
which relates the wavelength-integrated extinction 
to the total dust volume,
to place a lower limit on the dust depletion.
This, together with the observationally-derived
gas-phase abundances, allows us to rule out
the B-star, solar, and protosolar abundances
as the interstellar reference standard
and support the GCE-augmented protosolar abundances 
as a viable representation of the interstellar abundances.
%
%
\end{abstract}


\section{Introduction}\label{sec:intro}
Elements in the interstellar medium (ISM) 
exist in the form of gas or dust.
The interstellar gas-phase abundances of elements
can be measured from their optical and ultraviolet (UV)
spectroscopic absorption lines. The elements ``missing'' 
from the gas phase are bound up in dust grains,
known as ``interstellar depletion''. 
The dust-phase abundance of an element is 
derived by assuming a ``reference abundance''
and then from which subtracting off 
the gas-phase abundance. 
The reference abundance 
(also known as ``interstellar abundance'',  
or ``cosmic abundance'') of an element
is the total abundance of this element 
(both in gas and in dust).

The interstellar abundances of heavy, 
dust-forming elements such as C, O, Mg,
Si and Fe provide insight into the maximum 
amount of raw materials available for making 
the dust to account for the observed extinction, 
i.e., any dust models should not use more elements 
than what is available in the ISM.
This involves two unknown (or not well determined)
sets of abundances for the dust-forming elements: 
the interstellar reference abundances 
and the gas-phase abundances.
Although it is well recognized that
a viable dust model should satisfy  
the cosmic abundance constraints,
what might be the most appropriate 
set of interstellar reference abundances 
and what are the true gas-phase abundances 
of the dust-forming elements
have been subjects of much discussion 
in the past decades.
Historically, the interstellar abundances 
are commonly assumed to be solar. 
The gas-phase abundances of C and O
are often assumed to be invariable with 
respect to the local interstellar conditions
(e.g., see Cardelli et al.\ 1996, Meyer et al.\ 1998)
while Mg, Si and Fe appear to be fully depleted
from the gas phase 
(i.e., their gas-phase abundances are negligible).

In the late 1990s, it was argued that,
because of their young ages, the interstellar
abundances might be better represented by 
those of B stars and young F, G stars 
which are ``subsolar'' (i.e., $\simali$60--70\%  
of solar; Snow \& Witt 1995, 1996, Sofia \& Meyer 2001).
However, if the interstellar abundances 
are indeed ``subsolar'', the ISM seems to lack 
enough raw material to make the dust to account for 
the observed interstellar extinction (Li 2005).
Meanwhile, the published solar abundances have
undergone major changes over the years 
(see Table~\ref{tab:abund}).
The most recently determined solar abundances 
(Asplund et al.\ 2009; hereafter A09)  
are significantly reduced from their earlier values
(e.g., Anders \& Grevesse 1989).
It is interesting to note that,
using the non-local thermodynamic 
equilibrium (NLTE) techniques,
Przybilla et al.\ (2008) and Nieva \& Przybilla (2012) 
derived the photospheric abundances of heavy elements 
for unevolved early B-type stars.
They found that the photospheric abundances of 
those B stars are in close agreement 
with the A09 solar abundances
(see Table~\ref{tab:abund}).   

Lodders (2003) argued that the currently observed solar
photospheric abundances (relative to H) 
must be lower than those of the proto-Sun 
because helium and other heavy elements
have settled toward the solar interior 
since the time of its formation 
$\simali$4.55$\Gyr$ ago.
Lodders (2003) further suggested that 
the protosolar abundances derived from 
the photospheric abundances by considering 
settling effects are more representative 
of the solar system abundances.
With the settling effects taken into account,
the A09 solar abundances are roughly
consistent with the proto-Sun abundances
of Lodders (2003).
On the other hand, as the Galaxy evolves,
heavy elements are expected to be enriched
(Chiappini et al.\ 2003). 
Draine (2015) suggested that the protosolar abundances  
augmented by Galactic chemical enrichment (GCE) 
over the past 4.55\,Gyrs might be the best estimate 
for the interstellar abundances 
in the solar neighborhood.

Regarding the gas-phase abundances of 
the dust-forming elements C, O, Mg, Si and Fe, 
numerous observational studies carried out
in the past decade do not appear to favor
a constant gas-phase C/H abundance
of $\simali$140$\ppm$ (Cardelli et al.\ 1996)
and a constant gas-phase O/H abundance
of $\simali$320$\ppm$ (Meyer et al.\ 1998)
which were commonly adopted by dust models.
Also, contrary to what is commonly assumed by 
dust models, there seems to be nontrivial 
amounts of gas-phase Si, Mg and Fe 
(10\% or more in many sight lines; Jensen et al.\ 2010).
%

Cardelli et al.\ (1996) determined 
the gas-phase C abundance to be
$\cgas\approx140\pm20\ppm$
from the weak intersystem line of C\,II] 
at 2325$\Angstrom$ obtained with 
the Goddard High Resolution Spectrograph (GHRS)
on board the {\it Hubble Space Telescope} (HST)
for six sightlines which exhibit a wide range 
of extinction variation.
They found that $\cgas$ shows no dependence 
on the physical condition of the gas
(also see Sofia et al.\ 1997, 2004).
      %
      %
However, for the 21 sightlines studied by
Parvathi et al.\ (2012) and Sofia et al.\ (2011),
      $\cgas$ appears to decrease with
      $\langle\nH\rangle$, the mean density of H.
      This was based on 
      the strong 1334$\Angstrom$ C\,II 
      absorption line obtained with HST/STIS.
      These sight lines include 
      a variety of Galactic disk environments 
      characterized by different extinction
and sample paths ranging over three orders 
of magnitude in $\langle n_{\rm H}\rangle$.

Meyer et al.\ (1998) obtained high S/N ratio 
echelle spectra of the weak OI 1356$\Angstrom$
absorption of seven nearby diffuse clouds.
They derived a mean abundance 
      of $\ogas\approx319\pm14\ppm$ 
      and found no statistically 
      significant variations 
      among the sight lines and no evidence of 
      density-dependent O depletion. 
      %
In contrast, 
using the {\it Space Telescope Imaging Spectrograph}
(STIS) on board HST, Cartledge et al.\ (2001) performed
high-resolution observations 
of the OI 1356$\Angstrom$ absorption 
      of 11 translucent clouds and
      found an appreciably lower $\ogas$ 
      which suggests a trend toward an enhanced 
      O depletion for denser clouds.
Jensen et al.\ (2005) analyzed the HST/STIS 
and HST/GHRS spectra of
the OI 1356$\Angstrom$ absorption
of 10 sight lines and found
a trend of increasing O depletion 
with $\RV$ and $f({\rm H_2})$,
the fraction of H in molecular form.

Based on the HST/STIS echelle spectra
of the 1239, 1240$\Angstrom$ absorption of
Mg\,II, Cartledge et al.\ (2006) determined $\mggas$,
the gas-phase Mg abundance of 47 sight lines 
extending up to 6.5\,kpc through the Galactic disk 
which probe a variety of interstellar environments, 
covering ranges of $\simali$4 orders of magnitude 
in $f({\rm H_2})$ and over two orders of magnitude 
in $\langle n_{\rm H}\rangle$. 
They found that the depletion of Mg
is density-dependent. 

Miller et al.\ (2007) determined the gas-phase 
Si and Fe abundances based on the HST/STIS data
of the Si\,II] line at 2335$\Angstrom$
and the Fe\,II lines at 1142, 2234, 2249, 2260, 
and 2367$\Angstrom$ for six translucent clouds
which sample a variety of extinction characteristics 
as indicated by their $\RV$ values, 
which range from 2.6 to 5.8.\footnote{%
  $R_V\equiv A_V/E(B-V)$ is
  the total-to-selective extinction ratio,
  where $A_V$ is the visual extinction,
  $E(B-V)\equiv A_B-A_V$ is the reddening 
  or color excess between $A_V$ and $A_B$, 
  the $B$-band extinction.
  }
They found that $\sigas$ and $\fegas$ vary
from one sight line to another.
Similarly, Haris et al.\ (2016) derived
$\sigas$ for 131 sight lines based on
the HST/STIS, HST/GHRS and 
the {\it International Ultraviolet Explorer} (IUE) data 
and found that $\sigas$ is correlated with 
$\langle n_{\rm H}\rangle$ and $f({\rm H_2})$.
%

%

The interstellar extinction and elemental abundances 
provide crucial constraints on
the composition and size of interstellar dust.
Practically, most of the extinction modeling 
efforts have been so far directed to 
the Galactic average extinction curve
which is obtained by averaging over many 
clouds of different gas and dust properties
(e.g., see Mathis et al.\ 1977, 
Draine \& Lee 1984,
D\'esert et al.\ 1990, 
Mathis 1996,
Li \& Greenberg 1997, 
Weingartner \& Draine 2001,
Zubko et al.\ 2004,
Jones et al.\ 2013),
despite the fact that the interstellar extinction curves
actually exhibit considerable variations
from one sightline to another.
Also, most of the extinction modeling 
efforts have assumed a solar abundance
and a constant gas-phase abundance 
for the dust-forming elements,
even though as discussed above
the interstellar reference abundance
remains unknown and the gas-phase abundances
exhibit appreciable variations with the local
interstellar environments. 
Therefore, by modeling the Galactic 
average extinction curve, unavoidably, 
any details concerning 
the relationship between the dust properties 
and the physical and chemical conditions of 
the interstellar environments 
would have been lost.

In this work we will examine the extinction 
and elemental abundances of ten interstellar
lines of sight for each of which both the extinction curve
and the gas-phase abundances for all the major 
dust-forming elements C, O, Mg, Si and Fe have 
been observationally determined. 
Our goal is to investigate 
which would be the most appropriate
set of interstellar reference abundances, 
those of B stars (Przybilla et al.\ 2008), 
solar (A09), protosolar (Lodders 2003), 
or protosolar+GCE (Draine 2015)?
To reduce the number of model parameters,
we will not model the extinction curves,
instead, we will simply apply 
the model-independent 
Kramers-Kronig (KK) relation (Purcell 1969)
to relate the wavelength-integrated extinction
to the total dust volume.
Then, for each assumed set of reference abundances,
we will explore whether the remaining elements,
after subtracting off their gas-phase abundances,
are sufficient for accounting for the total dust volume
derived from the KK relation.
Our approach is essentially model-independent
since it does not require the knowledge 
of the detailed optical properties and
size distribution of the dust. 
All we need is a general assumption 
of the dust composition 
(e.g., silicate, graphite, oxides and iron).

This paper is organized as follows. 
We first compile in \S\ref{sec:sample}
a ``gold'' sample consisting of 
all the (10) sightlines for which 
the extinction curves from the near infrared (IR)
to the far UV and the gas-phase abundances of 
C, O, Mg, Si, and Fe have been
observationally determined.
Based on the measured gas-phase abundances 
of the dust-forming elements and the adopted
interstellar reference abundances, 
we derive in \S\ref{sec:Vdust} for each line of sight
the total dust volumes. We assume that interstellar dust
is made of (i) graphite and iron-containing amorphous silicate, 
(ii) graphite and iron-lacking amorphous silicate plus iron oxides,
or (iii) graphite and iron-lacking amorphous silicate plus iron.
We then apply the KK relation to determine in \S\ref{sec:KK}
the wavelength-integrated extinction from the total dust volume.
In \S\ref{sec:results} we compare the wavelength-integrated 
extinction derived from the KK relation with that derived from
observations
and find that only if the interstellar abundances
are like the GCE-augmented protosolar abundances would 
the KK-based wavelength-integrated extinction exceed
the observation-based wavelength-integrated extinction,
which is implied by the KK relation. 
These results as well as the ``missing O'' problem
and the relations between the extinction-to-gas ratios
and the interstellar physical and chemical conditions 
of the lines of sight in this ``gold'' sample
are also discussed in \S\ref{sec:results}.
Finally, we summarize our major results 
in \S\ref{sec:summary}.
%

%
%

\section{The Sample}\label{sec:sample}
We first search for in the literature 
as many interstellar sightlines as possible 
for which both the extinction curves have been 
determined from the near-IR to the far-UV and 
the gas-phase abundances have been measured 
for all the major dust-forming elements 
C, O, Mg, Si, and Fe.
To this end, we find ten such lines of sight
and tabulate in Table~\ref{tab:GasAbund} 
the gas-phase abundances of C, O, Mg, Si, and Fe
as well as the column densities of atomic hydrogen $\NHI$, 
molecular hydrogen $\NHH$ 
and the total hydrogen column densities 
$\NH=\NHI+2\NHH$.
The extinction parameters 
$c_{1}^{\prime}$, $c_{2}^{\prime}$,
$c_{3}^{\prime}$, $c_{4}^{\prime}$,
$\xo$ and $\gamma$ 
as well as $\AV$, $E(B-V)$ and $\RV$
are taken from Valencic et al.\ (2004)
and tabulated in Table~\ref{tab:extpara}.
These parameters characterize the UV extinction 
measured by the {\it International Ultraviolet Explorer} (IUE)
at $3.3 < \lambda^{-1} < 8.7\mum^{-1}$
as a sum of three components: 
a linear background,
a Drude profile for the 2175$\Angstrom$ extinction bump,
and a far-UV nonlinear rise 
at $\lambda^{-1} > 5.9\mum^{-1}$:
\begin{equation}\label{eq:A2AV1}
A_\lambda/A_V = c_1^{\prime} + c_2^{\prime}\,x 
              + c_3^{\prime}\,D(x,\gamma,\xo) 
              + c_4^{\prime}\,F(x) ~~~,
\end{equation}
\begin{equation}\label{eq:A2AV2}
D(x,\gamma,\xo) = \frac{x^2}
  {\left(x^2-\xo^2\right)^2 + x^2\gamma^2} ~~~,
\end{equation}
\begin{equation}\label{eq:A2AV3}
F(x) = \left\{\begin{array}{lr} 
0 ~, & x < 5.9\mum^{-1} ~~~,\\

0.5392\,\left(x-5.9\right)^2 
     + 0.05644\,\left(x-5.9\right)^3 ~, 
 & x \ge 5.9\mum^{-1} ~~~,\\
\end{array}\right.
\end{equation}
where $A_\lambda$ is the extinction 
at wavelength $\lambda$,
$x\equiv 1/\lambda$ 
is the inverse wavelength in $\mu$m$^{-1}$, 
$c_1^{\prime}$ and $c_2^{\prime}$ define 
the linear background, 
$c_3^{\prime}$ defines the strength of 
the 2175$\Angstrom$ extinction bump
which is approximated by $D(x,\gamma,\xo)$,
a Drude function which peaks at 
$\xo\approx4.6\mum^{-1}$ and has 
a FWHM of $\gamma$, 
and $c_4^{\prime}$ defines 
the nonlinear far-UV rise.\footnote{%
  This parametrization was originally introduced
   by Fitzpatrick \& Massa (1990; hereafter FM90)
   for the interstellar reddening
\begin{equation}\label{eq:E2A}
E(\lambda-V)/E(B-V) = R_V \left(A_\lambda/A_V - 1\right)
              = c_1 + c_2\,x 
              + c_3\,D(x,\gamma,\xo) 
              + c_4\,F(x) ~~~,
\end{equation}
where $E(\lambda-V) = A_\lambda - A_V$.
The extinction parameters of Valencic et al.\ (2004)
relate to the FM90 parameters through
\begin{equation}
c_j^{\prime} = \left\{\begin{array}{lr} 
c_j/R_V + 1 ~, & j=1 ~~~,\\
c_j/R_V ~, & j=2, 3, 4 ~~~.\\
\end{array}\right.
\end{equation}
}
In the following, we will refer to the parametrization 
described by Equations (\ref{eq:A2AV1}--\ref{eq:A2AV3})
as the FM parametrization.

For each line of sight, we aim at integrating 
the extinction over wavelength from 0 to $\infty$
(i.e., $\int_{0}^{\infty} A_\lambda\,d\lambda$).
For practical reasons, this is not possible 
since the extinction curve is observationally 
determined only over a limited wavelength range, 
usually from the near-IR to the far-UV.
We shall therefore ``construct'', for each sightline,
the extinction curve from 912$\Angstrom$ to 1$\cm$
and then obtain 
\begin{equation}\label{eq:Aintobs}
\Aintobs \equiv \int_{\rm 912\Angstrom}^{1\cm} 
A_\lambda\,d\lambda ~~.
\end{equation}

\begin{figure*}
	\centering	
	\includegraphics[width=0.5\textwidth,height=1.0\hsize]{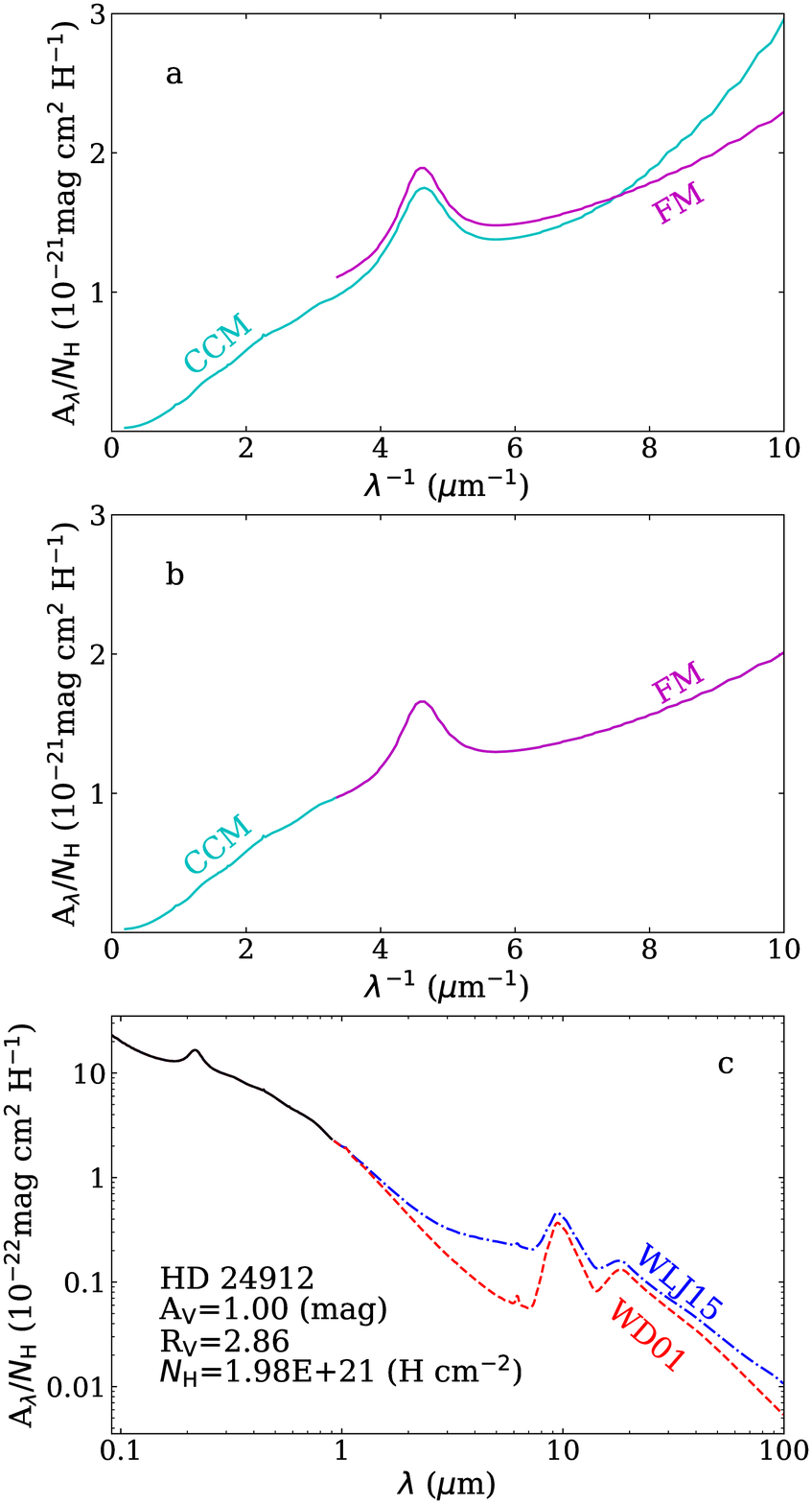}
	\caption{\footnotesize
        \label{fig:extconstruct}
         Constructing the interstellar extinction curve
         from the far-UV to the far-IR
         for the line of sight toward HD\,24912.
         Panel (a): Comparison of the FM extinction curve
         at $\lambda^{-1}>3.3\mum^{-1}$ (purple line)
         and the CCM curve from the near-IR to far-UV (cyan line).
         The FM curve represents the observed IUE extinction curve,
         while the CCM curve is derived from the CCM parameterization
         with $\RV=2.86$ for HD\,24912.
         Panel (b): The FM curve is multiplied by a factor to smoothly    
         join the CCM curve at $\lambda^{-1}=3.3\mum^{-1}$.
         Panel (c): The interstellar extinction curve 
         from the far-UV to the far-IR, 
         with the FM curve for $\lambda^{-1}>3.3\mum^{-1}$,
         the CCM curve for  $1.1\mum^{-1} < \lambda^{-1} < 3.3\mum^{-1}$,
         and the $\RV=3.1$ model curves of 
         Weingartner \& Draine (2001; red dashed line) 
         and Wang, Li \& Jiang (2015a; blue dot-dashed line) 
         for $0.9\mum < \lambda < 1\cm$.
         }
\end{figure*}

\begin{figure*}
\centering	
\includegraphics[width=0.5\textwidth,height=1.0\hsize]{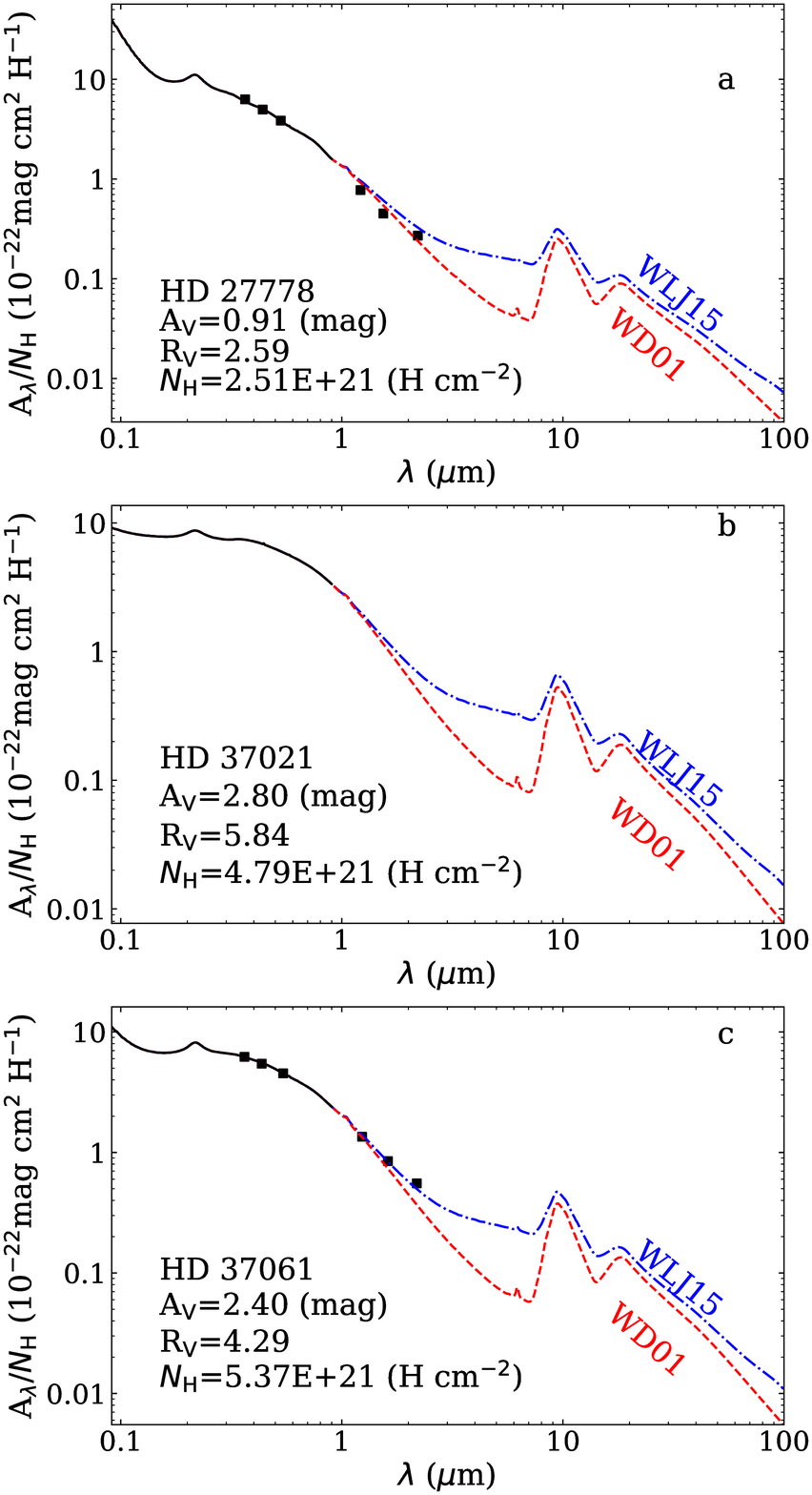}
\caption{\footnotesize
         \label{fig:curve1}
         The interstellar extinction curves 
         from the far-UV to the far-IR
         for the lines of sight toward 
         HD\,27778 (a), HD\,37021 (b) and HD\,37061 (c),
         with the FM curve for $\lambda^{-1}>3.3\mum^{-1}$,
         the CCM curve for  $1.1\mum^{-1} < \lambda^{-1} < 3.3\mum^{-1}$,
         and the $\RV=3.1$ model curves of 
         Weingartner \& Draine (2001; red dashed line) 
         and Wang, Li \& Jiang (2015a; blue dot-dashed line) 
         for $0.9\mum < \lambda < 1\cm$.
         Whenever available, broadband photometric 
         extinction data (see Table~\ref{tab:extpara})
         are shown as black squares.
         The U, B, V, J, H, K extinction data
         are superimposed on the extinction curves
         of  HD\,27778 (a) and HD\,37061 (c).
         }
\end{figure*}

\begin{figure*}
\centering	
\includegraphics[width=0.5\textwidth,height=1.0\hsize]{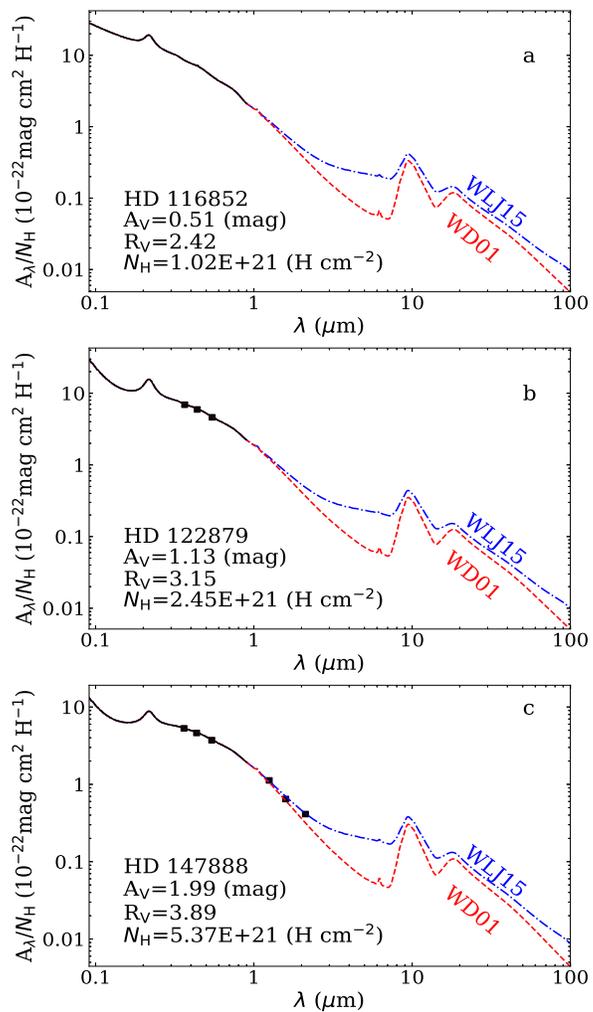}
\caption{\footnotesize
         \label{fig:curve2}
         Same as Figure~\ref{fig:curve1}
         but for HD\,116852, HD\,122879 and HD\,147888.
         Black squares show the U, B, V extinction data 
         for HD\,122879 and the U, B, V, J, H, K extinction 
         data for HD\,147888.
         }
\end{figure*}

\begin{figure*}
	\centering	
	\includegraphics[width=0.5\textwidth,height=1.0\hsize]{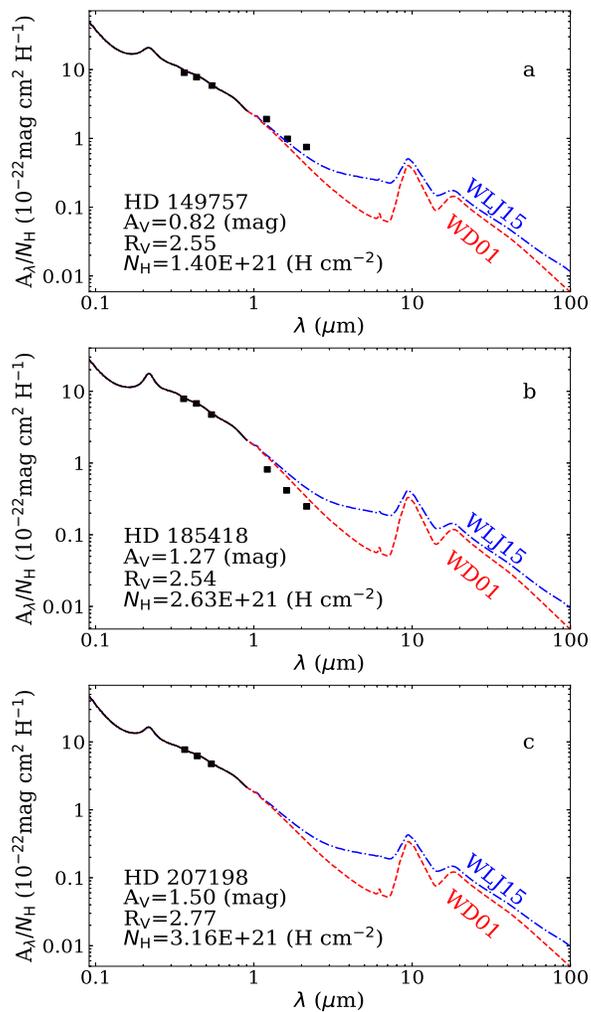}
	\caption{\footnotesize
		\label{fig:curve3}
         Same as Figure~\ref{fig:curve1}
         but for HD\,149757, HD\,185418 and HD\,207198.
         Black squares show the U, B, V, J, H, K extinction 
         data for HD\,149757 and HD\,185418 
         and the U, B, V extinction data for HD\,207198.
         }
\end{figure*}

We construct the extinction curves as follows
(see Figure~\ref{fig:extconstruct}).
For $3.3 < \lambda^{-1} < 11\mum^{-1}$,
we represent the extinction by Equation (\ref{eq:A2AV1})
with the extinction parameters taken from
Table~\ref{tab:extpara}.\footnote{%
   We note that, although Equation (\ref{eq:A2AV1})
   was originally derived from the IUE data 
   over $3.3< \lambda^{-1} < 8.7\mum^{-1}$,
   Gordon et al.\ (2009) found that 
   the general shapes of the extinction curves
   at $8.4< \lambda^{-1} < 11\mum^{-1}$ obtained 
   by the {\it Far Ultraviolet Spectroscopic Explorer} 
   (FUSE) are broadly consistent with extrapolations 
   from the IUE extinction curves. 
   }
For $1.1 < \lambda^{-1} < 3.3\mum^{-1}$,
we compute the extinction from the parametrization 
of Cardelli et al.\ (1989; hereafter CCM).
The CCM parametrization involves only one parameter,
that this, $\RV$.
As illustrated in Figure~\ref{fig:extconstruct}a,
there is often a discontinuity 
between the FM parametrization 
at $\lambda^{-1} > 3.3\mum^{-1}$
and the CCM parametrization 
at $\lambda^{-1} < 3.3\mum^{-1}$.
To comply with the observed extinction-to-gas ratio
$\AV/\NH$, we multiply the FM extinction curve
by a factor to smoothly join the CCM curve
(see Figure~\ref{fig:extconstruct}b).
For $0.9\mum < \lambda < 1\cm$,
we approximate the extinction 
by the model extinction calculated from
the standard silicate-graphite-PAH model
of Weingartner \& Draine (WD01) for $\RV=3.1$
(see Figure~\ref{fig:extconstruct}c).
Note that the WD01 model extinction curve
exhibits a deep minimum at $\simali$5--8$\mum$,
whereas numerous observations made with
the {\it Infrared Space Observatory} (ISO)
and the {\it Spitzer Space Telescope}
have shown that the mid-IR extinction 
at $3\mum <\lambda< 8\mum$
is flat for both diffuse and dense environments
(Lutz 1999, Indebetouw et al.\ 2005,
Jiang et al.\ 2006, 
Flaherty et al.\ 2007,
Gao et al.\ 2009,
Nishiyama et al.\ 2009, 
Wang et al.\ 2013,
Xue et al.\ 2016,
Hensley \& Draine 2020a).
By introducing a population of
very large, micron-sized graphitic grains,
Wang et al.\ (2015a; hereafter WLJ15) closely 
reproduced the observed flat mid-IR extinction.
Therefore, for $0.9\mum < \lambda < 1\cm$,
we will also approximate the extinction 
by the WLJ15 model extinction
(see Figure~\ref{fig:extconstruct}c).
As a result, for each sightline we derive
two extinction curves (which we refer to
as ``WD01'' and ``WLJ15''; 
see Figures~\ref{fig:curve1}--\ref{fig:curve3})
and integrate the extinction (per hydrogen column) 
over 912$\Angstrom$ and 1$\cm$ to obtain 
$\AintobsWD/\NH$ and $\AintobsWLJ/\NH$.
We have also tried our best to compile 
for each line of sight
the broadband photometric extinction data
(see Table~\ref{tab:extpara}).
Whenever available, they are displayed
as black squares
superimposed on the synthesized extinction curves.
As shown in Figures~\ref{fig:curve1}--\ref{fig:curve3},
the synthesized extinction curves of all lines of sight
closely agree with the observationally-determined 
U, B and V extinction. While the WD01 curve of HD\,27778 
and the WLJ15 curves of HD\,37061 and HD\,147888
agree with their J, H, and K extinction data,
the WD01 curve of HD\,185418 is somewhat higher
and the WLJ15 curve of HD\,149757 is somewhat lower
than their J, H, and K extinction data.

\section{Total Dust Volumes as Constrained 
            by the Elemental Abundances
            }\label{sec:Vdust}
For an assumed dust composition, for each line of sight 
we can estimate the total dust volume per H nucleon 
($\Vdust$/H) from the adopted set of interstellar reference abundances
and the observationally-determined gas-phase abundances.
For the dust composition, we will first consider a mixture of
amorphous silicate and graphite. It is well recognized
that amorphous silicate is a ubiquitous component of the Universe
as revealed by the 9.7$\mum$ Si--O stretching feature 
and the 18$\mum$ O--Si--O bending feature 
seen either in absorption or in emission (see Henning 2010).
There must also be a population of carbon dust, 
although its exact composition remains unknown 
(see Henning \& Salama 1998).
This is because, as discussed in \S\ref{sec:intro},
C is partially depleted from the gas phase 
and silicate alone is not sufficient to account for
the observed extinction (see Mishra \& Li 2015, 2017).
In this work we assume that all the C atoms missing
from the gas phase are locked up in graphite grains 
since presolar graphite grains have been identified 
in primitive meteorites and the interstellar 2175$\Angstrom$
extinction bump, the strongest interstellar absorption feature,
is generally attributed to small graphitic grains. 

Let $\cism$ be the interstellar C abundance (relative to H).
We calculate the volume per H nucleon ($\Vgra$/H)
occupied by graphite dust from
%
%
\begin{equation}\label{eq:Vgra2H}
\frac{\Vgra}{\rm H} = \left\{\cism-\cgas\right\}
         \frac{\muc \mH}{\rhogra} ~~,
\end{equation}
where $\muc=12$ is the atomic weight of C, 
$\rho_{\rm gra}\approx2.2\g\cm^{-3}$
is the mass density of graphite, 
and $\mH=1.66\times10^{-24}\g$
is the mass of a hydrogen atom.
For the silicate component, 
we assume an even mixture of
pyroxene (Mg$_x$Fe$_{1-x}$SiO$_3$) 
and olivine (Mg$_{2x}$Fe$_{2-2x}$SiO$_4$) 
compositions, where $0\simlt x\simlt 1$.
Therefore, we assign 3.5 O atoms for each Si atom.
We calculate the volume per H nucleon ($\Vsil$/H)
occupied by silicate dust from
\begin{equation}\label{eq:Vsil2Ha}
\begin{aligned}
\frac{\Vsil}{\rm H} &= \Bigg\{
	\left(\mgism-\mggas\right) \mumg +
        \left(\feism-\fegas\right) \mufe \\
        & + \left(\siism-\sigas\right) \musi +
       3.5 \left(\siism-\sigas\right) \muo 
         \Bigg\}
         \frac{\mH}{\rhosil} ~~,
\end{aligned}
\end{equation}
where $\mgism$, $\feism$, $\siism$
and $\oism$ are respectively the interstellar
Mg, Fe, Si and O abundances (relative to H), 
$\mumg$, $\mufe$, $\musi$ and $\muo$ 
are respectively the atomic weights of Mg, Fe, Si and O,
and $\rhosil\approx3.5\g\cm^{-3}$ is the mass density
of silicate dust. 

We consider four sets of interstellar reference abundances,
by adopting the abundances of 
B stars (Przybilla et al.\ 2008),
solar abundances (A09),
protosolar abundances (Lodders 2003), and
GCE-augmented protosolar abundances (Draine 2015)
as the interstellar abundances.
For each adopted set of interstellar abundances,
we calculate $\Vgra$/H and $\Vsil$/H 
from Equations (\ref{eq:Vgra2H},\ref{eq:Vsil2Ha})
for each sightline and tabulate the results
in Tables~\ref{tab:Bstar}--\ref{tab:protosolarGCE}.

So far, we have assumed that all the Fe atoms 
missing from the gas phase are depleted 
in amorphous silicate grains.
In the Galactic ISM, typically 90\% or more 
of the Fe is missing from the gas phase (Jenkins 2009), 
suggesting that Fe is the largest elemental
contributor to the interstellar dust mass 
after O and C and accounts for $\simali$25\% 
of the dust mass in diffuse interstellar regions.
However, as yet we know little about the nature 
of the Fe-containing material.
Silicate grains provide a possible reservoir 
for the Fe in the form of interstellar pyroxene
or olivine 
analogues. 
Nevertheless,
iron abundances and depletions in the ISM 
often diverge from the pattern shown by Si and Mg,
suggesting that Fe is not tied to the same grains 
as Si, and therefore most silicate grains
are likely Mg-based. 
Also, the shape and strength of the interstellar
9.7$\mum$ silicate absorption feature suggest that 
the silicate material is Mg-rich rather than 
Fe-rich (Poteet et al.\ 2015) and hence
a substantial fraction ($\simali$70\%) of 
the interstellar Fe might be in other forms 
such as iron oxides, iron sulfides,\footnote{%
      S is abundant in the ISM 
      and previous studies have suggested 
      that S is not depleted from the gas-phase.
      However, White \& Sofia (2011) analyzed 
      the strong S\,II 
      1250, 1253, 1259$\Angstrom$ lines 
      of 28 sight lines obtained with HST/STIS
      and found S is depleted into grains. 
      Westphal et al.\ (2019) found that 
      the Fe L-edge absorption spectrum of 
      continuum X-rays from Cygnus X-1 
      is consistent with the hypothesis that 
      Fe is sequestered principally in metals, 
      with Fe evenly divided between sulfide and metal
      and less than $\simali$38.5\% (2$\sigma$)
      of Fe residing in amorphous silicates.
      In this work, we will neglect S-bearing grains
      (e.g., iron sulfides) since all the Fe atoms have
      already been included in our model calculations
      and whether they reside in sulfides or oxides 
      would not increase much the total dust volume.
      }
or metallic iron (see Draine \& Hensley 2013, 
Dwek 2016, Hensley \& Draine 2017).\footnote{%
    Rogantini et al.\ (2020) recently analyzed
    the X-ray spectroscopy of the magnesium and the silicon 
    K-edges for eight bright X-ray binaries,
    distributed in the neighbourhood of the Galactic center, 
    detected with the High Energy Transmission Grating 
    Spectrometer (HETGS) aboard the {\it Chandra X-Ray Telescope}. 
    However, they found that amorphous olivine with a composition 
    of MgFeSiO$_4$ is the most representative compound 
    along all these lines of sight,  at least in the diffuse ISM 
    in the inner regions of the Milky Way 
    within $\simali$5\,kpc of the center.
    %
    }
Therefore, we will also consider the ISM 
to consist of graphite, Fe-lacking silicates
(i.e., forsterite Mg$_2$SiO$_4$ and enstatite MgSiO$_3$),
and three types of iron oxides 
(i.e., w\"ustite FeO, 
haematite Fe$_2$O$_3$, 
and magnetite Fe$_3$O$_4$).\footnote{%
   If Fe is tied up in iron carbides 
  (e.g., Fe$_3$C; see Nuth et al.\ 1985), 
  its contribution to the wavelength integral 
  of extinction would be smaller than that of
  iron oxides. This is because, in comparison 
  with iron oxides (at least  Fe$_3$O$_4$), 
  Fe$_3$C has a higher mass density of 
  $\rho\approx7.7\g\cm^{-3}$
  (and therefore a smaller grain volume)
  and a lower static dielectric constant
  $\varepsilon_0$ (and therefore a smaller
  $F(a/b,\varepsilon_0)$ factor;
  see Equation [\ref{eq:Fdef}]).
  Moreover, iron carbides consume C 
  which would have been locked up in carbon dust,
  while iron oxides, in addition to silicates,
  consume O which otherwise would be in the gas phase.
  Also, the 30$\mum$ emission feature 
  expected for Fe$_3$C (Forrest et al.\ 1981; 
  but also see Nuth et al.\ 1985)
  is not seen in the ISM.
  Therefore, we assume that Fe ties up in 
  iron oxides instead of iron carbides. 
  }
We assume that the Fe atoms missing from the gas phase
are evenly tied up in FeO, Fe$_2$O$_3$, and Fe$_3$O$_4$. 
In this case, we calculate the volumes of silicate dust
and iron oxides from
\begin{equation}\label{eq:Vsil2Hb}
\begin{aligned}
\frac{\Vsil}{\rm H} & = \Bigg\{
	\left(\mgism-\mggas\right) \mumg 
         + \left(\siism-\sigas\right) \musi \\
         & + 3.5 \left(\siism-\sigas\right) \muo 
         \Bigg\}
         \frac{\mH}{\rhosil^\prime} ~~,
\end{aligned}
\end{equation}
\begin{equation}\label{eq:VFeOa}
\begin{aligned}
\frac{\VFeOa}{\rm H} & = \frac{1}{3} \Bigg\{
	\left(\feism-\fegas\right) \mufe
        + \left(\feism-\fegas\right) \muo 
         \Bigg\}
         \frac{\mH}{\rhoFeOa} ~~,
\end{aligned}
\end{equation}
\begin{equation}\label{eq:VFeOb}
\begin{aligned}
\frac{\VFeOb}{\rm H} & = \frac{1}{3} \Bigg\{
	\left(\feism-\fegas\right) \mufe
        + \frac{3}{2} \left(\feism-\fegas\right) \muo 
         \Bigg\}
         \frac{\mH}{\rhoFeOb} ~~,
\end{aligned}
\end{equation}
\begin{equation}\label{eq:VFeOc}
\begin{aligned}
\frac{\VFeOc}{\rm H} & = \frac{1}{3} \Bigg\{
	\left(\feism-\fegas\right) \mufe
        + \frac{4}{3} \left(\feism-\fegas\right) \muo 
         \Bigg\}
         \frac{\mH}{\rhoFeOc} ~~,
\end{aligned}
\end{equation}
where $\rhosil^\prime\approx3.2\g\cm^{-3}$
is the mass density of Fe-lacking silicate,
$\rhoFeOa\approx5.7\g\cm^{-3}$, 
$\rhoFeOb\approx5.3\g\cm^{-3}$ and 
$\rhoFeOc\approx5.2\g\cm^{-3}$
are respectively the mass densities of
FeO, Fe$_2$O$_3$ and Fe$_3$O$_4$
(see Table~\ref{tab:DustMaterials}).
Again, for each adopted set of interstellar abundances,
we calculate $\Vgra$/H, $\Vsil$/H, $\VFeOa$,
$\VFeOb$ and $\VFeOc$ from 
Equations (\ref{eq:Vgra2H},\,\ref{eq:Vsil2Hb}--\ref{eq:VFeOc})
for each sightline and tabulate the results in 
Tables~\ref{tab:Bstar}--\ref{tab:protosolarGCE}.

Finally, we also consider the case of locking 
up all the Fe atoms missing from the gas phase 
in iron grains. In this case, the silicate dust volume
is the same as Equation (\ref{eq:Vsil2Hb}) and 
the volume of iron dust is
\begin{equation}\label{eq:VFe}
\begin{aligned}
\frac{\VFe}{\rm H} & = \Bigg\{
	\left(\feism-\fegas\right) \mufe
         \Bigg\}
         \frac{\mH}{\rhoFe} ~~,
\end{aligned}
\end{equation}
where $\rhoFe\approx7.8\g\cm^{-3}$
is the mass density of iron.
In Tables~\ref{tab:Bstar}--\ref{tab:protosolarGCE}
we also tabulate $\Vgra$/H, $\Vsil$/H and $\VFe$/H
calculated from 
Equations (\ref{eq:Vgra2H},\ref{eq:Vsil2Hb},\ref{eq:VFe})
respectively for four sets of interstellar reference abundances.

\section{Total Wavelength-Integrated Extinction
            as Constrained by the Total Dust Volume 
           }\label{sec:KK}
Let $\AintKK\equiv \int_{0}^{\infty} A_\lambda\,d\lambda$
be the extinction integrated over all wavelengths.
If, in the ISM, there are $N$ different types of dust species 
and $V_j$/H is the volume (per H nucleon) of the $j$-th dust type,
the KK relation relates 
the wavelength-integrated extinction
to the total volume (per H nucleon)  
occupied by dust through 
\begin{equation}\label{eq:kk}
\AintKK/\NH
= \int_{0}^{\infty} A_\lambda/\NH\,d\lambda 
= 1.086\times 3 \pi^2 
\sum_{j=1}^{N} F_j\,\left(V_j/{\rm H}\right) ~~,
\end{equation}
where the dimensionless factor $F_j$ is 
the orientationally-averaged polarizability 
of the $j$-th dust type relative to the polarizability 
of an equal-volume sphere, 
depending only upon the grain shape and 
the static (zero-frequency) dielectric constant 
$\varepsilon_0$ of the grain material (Purcell 1969).
In Table~\ref{tab:DustMaterials} we tabulate 
the $\varepsilon_0$ values for all the dust materials
of interest in this work (i.e., graphite, iron, 
FeO, Fe$_2$O$_3$, Fe$_3$O$_4$, 
MgFeSiO$_4$, Mg$_2$SiO$_4$, and MgSiO$_3$).

To calculate the $F$ factors,
we take the dust to be spheroids
with semiaxes $a,b,b$ 
(prolate if $a/b>1$, oblate if $a/b<1$),
then $F$ is related to the static dielectric 
constant $\varepsilon_0$ 
and the ``depolarization factors'' 
$L_a$ and $L_b=(1-L_a)/2$ through
\beq
F(a/b,\varepsilon_0) \equiv \frac{(\varepsilon_0-1)}{9}
\left[
\frac{1}{(\varepsilon_0-1)L_a + 1} + \frac{2}{(\varepsilon_0-1)L_b + 1}
\right]
~~~,
\label{eq:Fdef}
\eeq
where for prolates 
\beq
L_a = \frac{1-e^2}{e^2}
\left[\frac{1}{2e}\ln\left(\frac{1+e}{1-e}\right)-1\right]
~~~,
\eeq
and for oblates 
\beq
L_a = \frac{1+e^2}{e^2}\left[1-\frac{1}{e}\arctan(e)\right]
~~~,
\eeq
where
\beq
e\equiv |1-(b/a)^2|^{1/2} ~~~. 
\label{eq:e}
\eeq

\begin{figure}
\includegraphics[width=0.8\textwidth]{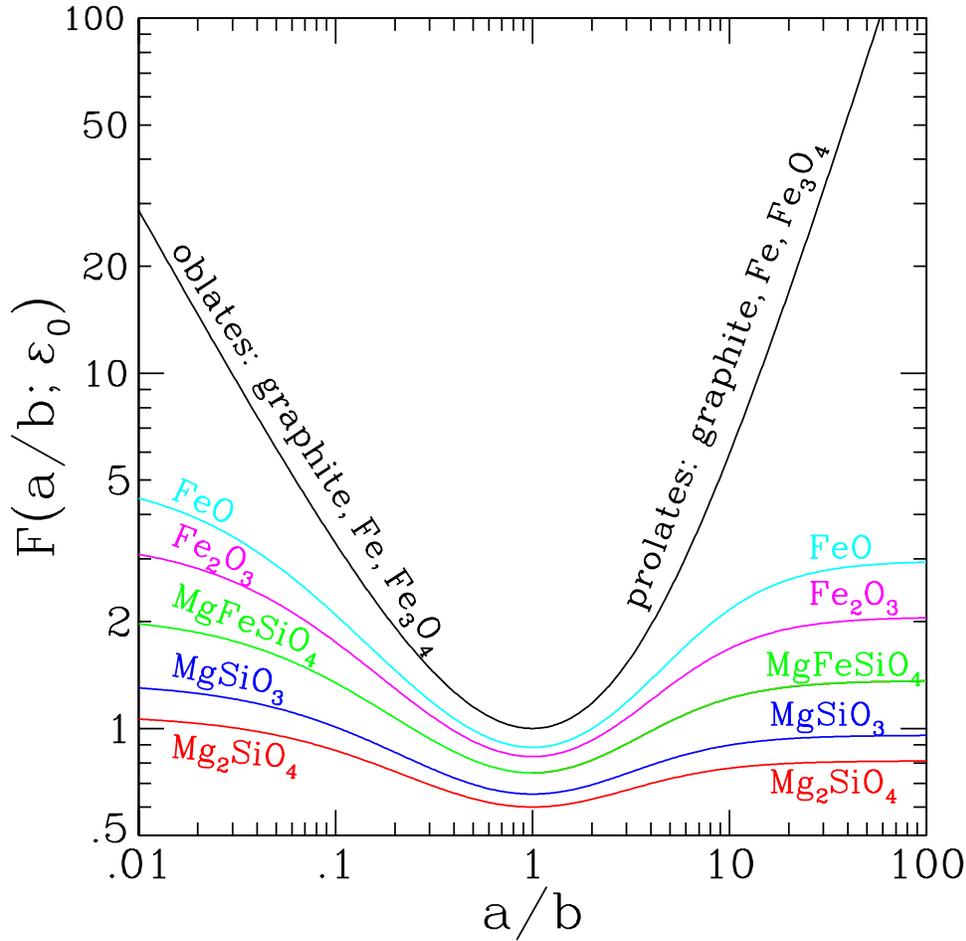}

\caption{
        \label{fig:dpol}
        \footnotesize
        $F(\varepsilon_0;{\rm shape})$ factors
        as a function of the axial ratio $a/b$
        for Mg$_2$SiO$_4$ (red line), 
        MgSiO$_3$ (blue line), 
        MgFeSiO$_4$ (green line),  
        FeO (cyan line), Fe$_2$O$_3$ (magenta line),
        as well as Fe, Fe$_3$O$_4$  and graphite (black line)
        of which the static dielectric constants
        are approximately 
        $\varepsilon_0$\,$\approx$\,5.5, 6.7, 10, 24,16, 
        $\infty$, $\infty$ and $\infty$, respectively (Li 2005).
        The grains are taken to be spheroidal
        with $a$ and $b$ being the semiaxis 
        along and perpendicular to the symmetry axis
        of the spheroid, respectively.
        The oblate dust is with axial $a/b<1$ 
        while the prolate is with axial $a/b>1$.
        For modestly elongated ($a/b\simlt 3$) 
        or flattened ($b/a\simlt 3$) silicate dust,
        the $\rm F$ factor is always smaller than unity. 
        }
\end{figure}

By making use of the $\varepsilon_0$ values 
shown in Table~\ref{tab:DustMaterials},
we calculate the $F$ factors as a function 
of grain shape (i.e., elongation $a/b$) 
for all the dust species considered here 
(see Figure~\ref{fig:dpol}).
It is apparent that for both dielectric materials
(e.g., MgFeSiO$_4$, Mg$_2$SiO$_4$, 
MgSiO$_3$, FeO, Fe$_2$O$_3$)
and conducting materials
(e.g., graphite, iron, Fe$_3$O$_4$),
the $F$ factors of modestly elongated 
or flattened grains do not deviate much
from unity. In this work, for each dust type
we will adopt the mean $F$ value
averaged over that calculated 
for $a/b=3$ prolates and $a/b=1/2$ oblates
(see Table~\ref{tab:DustMaterials}).
The $a/b=3$ prolate shape is chosen
because Greenberg \& Li (1996) found that
the 9.7 and 18$\mum$ silicate polarization 
toward the Becklin-Neugebauer (BN) object is 
best reproduced by $a/b=3$ core-mantle spheroids,
while the $a/b=1/2$ oblate shape is selected
because Lee \& Draine (1985) found that
the 3.1$\mum$ ice polarization of the BN object
is best explained by $a/b=1/2$ oblates.
Also, Hildebrand \& Dragovan (1995) found
that $a/b=1/2$ bare silicate oblates
could fit the 9.7$\mum$ polarization
of the BN object.

For each adopted set of interstellar reference abundances,
depending on how the Fe atoms missing from the gas phase
are divided among amorphous silicates, oxides and iron grains,
we have accordingly calculated in \S\ref{sec:Vdust}
the possible total dust volumes for each line of sight 
(see Tables~\ref{tab:Bstar}--\ref{tab:protosolarGCE}).  
Combining the dust volumes derived in \S\ref{sec:Vdust}
and the $F$ factors derived in this section, 
we calculate and tabulate in 
Tables~\ref{tab:Bstar}--\ref{tab:protosolarGCE}  
for each sightline $\AintKK$, 
the extinction integrated over all wavelengths, 
for each of the four interstellar reference abundances
(i.e., B-star, solar, protosolar, and
GCE-augmented protosolar abundances)
and each of three different dust mixtures
(i.e., graphite\,+\,Fe-containg silicate,
graphite\,+\,Fe-lacking silicate\,+\,iron oxides,
and graphite\,+\,Fe-lacking silicate\,+\,Fe).

It is interesting to note that the $\AintKK$ values
calculated for different dust mixtures are rather close.
Compared with the graphite\,+\,Fe-containg silicate mixture,
the graphite\,+\,Fe-lacking silicate\,+\,iron oxides mixture
contains more mass (on a per H nucleon basis). 
But the total dust volume of the latter exceeds 
that of the former by only several percent
(because of the mass-density differences
between Fe-containg silicates with iron oxides 
and Fe-lacking silicates).
Although the $F$ factors of Fe-containing silicates
are smaller than that of iron oxdies, they are larger
than that of Fe-lacking silicates. 
Adding these effects together, 
the resulting $\AintKK$ values of 
the graphite\,+\,Fe-lacking silicate\,+\,iron oxides mixture
differ little from that of 
the graphite\,+\,Fe-containing silicate mixture.
On the other hand, because of the higher mass density
of iron (compared with silicates), the total dust volume 
of the graphite\,+\,Fe-lacking silicate\,+\,Fe mixture
is smaller than that of 
the graphite\,+\,Fe-containg silicate mixture.
The higher $F$ factor of iron does not sufficiently
compensate the smaller dust volume and therefore
the graphite\,+\,Fe-lacking silicate\,+\,Fe mixture
actually has a lower $\AintKK$ value 
than the graphite\,+\,Fe-containg silicate mixture.

The aforementioned results are derived from
the KK-based wavelength-integrated extinction
of Purcell (1969) who assumed nonmagnetic grains
(see Equation [\ref{eq:kk}]).
For magnetic grains, incident electromagnetic waves 
certainly excite the oscillation of magnetic moments 
and result in the loss of the incident electromagnetic waves.
Therefore, the absorption due to magnetic dipole moments 
from ferromagnetic (Fe) and ferrimagnetic 
(Fe$_2$O$_3$ and Fe$_3$O$_4$) grains
could be appreciable and the wavelength integral
of extinction becomes
\begin{equation}\label{eq:kkmag}
\int_{0}^{\infty} A_\lambda/\NH\,d\lambda 
= 1.086\times 3 \pi^2 
\sum_{j=1}^{N} 
\left\{
F_j(a/b,\varepsilon_0) + F_j(a/b,\mu_0) 
\right\}
\left(V_j/{\rm H}\right) ~~,
\end{equation}
where the dimensionless factor
$F(a/b,\varepsilon_0)$, 
as defined in Equation (\ref{eq:Fdef}),
measures the electric response,
while the dimensionless factor
$F(a/b,\mu_0)$ 
measures the magnetic response
(see Draine \& Lazarian 1999):
\beq
F(a/b,\mu_0) \equiv \frac{(\mu_0-1)}{9}
\left[
\frac{1}{(\mu_0-1)L_a + 1} + \frac{2}{(\mu_0-1)L_b + 1}
\right]
~~~,
\label{eq:Fmagdef}
\eeq
where $\mu_0$ is the static (zero-frequency)
magnetic permeability of the grain material
(for nonmagnetic grains $\mu_0\approx1$
and therefore $F(a/b,\mu_0) \approx0$).
For grain materials having $\varepsilon_0\simgt3$
and $\mu_0\simgt3$, 
$F(a/b,\mu_0)\approx F(a/b,\varepsilon_0)$,
and we see from Equations (\ref{eq:kk}) 
and (\ref{eq:kkmag})
that the wavelength integral of extinction 
for magnetic grains (e.g., Fe, Fe$_2$O$_3$ and Fe$_3$O$_4$) 
would be increased by a factor of
$\left\{1+F(a/b,\mu_0)/F(a/b,\varepsilon_0)\right\}\approx2$
than it would have been 
had the grains been nonmagnetic. 
However, as demonstrated in Draine \& Lazarian (1999)
and Draine \& Hensley (2013), the magnetic dipole absorption
mostly occurs at $\lambda\simgt3\mm$
and its contribution to the extinction of interest here
is not important since we are mostly concerned 
with the UV, optical, near- and mid-IR interstellar extinction.
This justifies the neglect of the magnetic dipole
absorption of magnetic grains 
(i.e.,  Fe, Fe$_2$O$_3$ and Fe$_3$O$_4$) 
in this work.

\section{Results and Discussion}\label{sec:results}
For a fixed amount of dust materials,
$\AintKK/\NH$ gives the {\it maximum} possible 
amount of wavelength-integrated extinction 
per H nucleon (see Equation [\ref{eq:kk}]), 
while both $\AintobsWD/\NH$ and $\AintobsWLJ/\NH$
are obtained by integrating the ``observed'' extinction
over a {\it finite} wavelength range 
(see Equation [\ref{eq:Aintobs}]).
For an interstellar reference abundance standard
to be a {\it viable} representation of the ``true'' 
interstellar abundances, the amounts of 
dust-forming elements available for making dust 
have to be {\it sufficient} to account for 
the observed extinction.
This implies that $\AintKK/\NH$ should exceed
$\AintobsWD/\NH$ and $\AintobsWLJ/\NH$
since the extinction $A_\lambda/\NH$ is always 
positive and therefore it is always true that
$\int_{0}^{\infty} A_\lambda/\NH\,d\lambda\,>\,
\int_{912\Angstrom}^{1\cm} A_\lambda/\NH\,d\lambda$.
For an adopted set of interstellar reference abundances,
if the corresponding $\AintKK/\NH$ is smaller than 
or equal to $\AintobsWD/\NH$ and $\AintobsWLJ/\NH$,
it simply means that the adopted reference abundances
are not viable since there would be insufficient amounts 
of dust-forming elements to make the dust to
cause the observed amounts of extinction.

\begin{figure*}
	\centering	
	\vspace{-10mm}
	\includegraphics[width=1.0\textwidth,height=0.9\textheight]{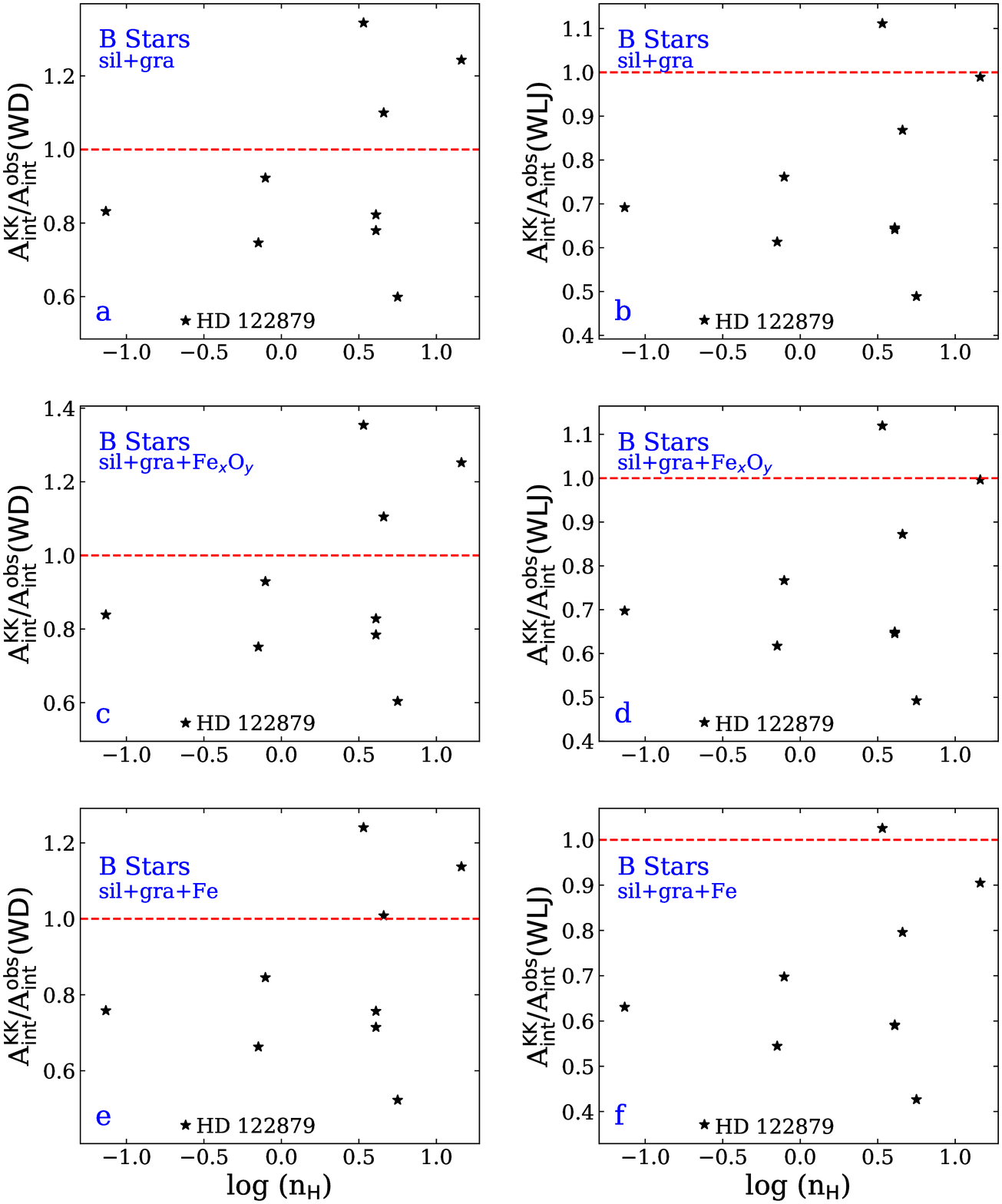}
	\vspace{-10mm}
	\caption{\footnotesize
	\label{fig:Bstar}
        Comparison of the KK-based wavelength-integrated 
        extinction $\AintKK$ with the observation-based
        wavelength-integrated extinction 
        $\AintobsWD$ and $\AintobsWLJ$
        for the ``gold'' sample of 10 lines of sight,
        with $\AintKK$ derived from the assumption
        of the abundances of B stars as the interstellar
        reference abundances and the dust as mixtures 
        of (i) graphite and Fe-containing amorphous silicates (a, b),
        (ii) graphite and Fe-lacking amorphous silicates 
        plus iron oxides (c, d), and (iii) graphite and 
        Fe-lacking amorphous silicates plus iron (e, f).
        The observation-based wavelength-integrated 
        extinction $\AintobsWD$ and $\AintobsWLJ$
        are obtained by integrating the ``observed'' 
        extinction curve from 912$\Angstrom$ to 1$\cm$,
        with the extinction curve at $0.9\mum < \lambda < 1\cm$
        respectively approximated by
        the $\RV=3.1$ model curve of 
        Weingartner \& Draine (2001)
        and of Wang, Li \& Jiang (2015a).
        For the adopted interstellar reference standard
        to be viable, we require for all sightlines
        $\AintKK/\AintobsWD >1$
        and $\AintKK/\AintobsWLJ > 1$.
	}
\end{figure*}

\begin{figure*}
	\centering	
	\includegraphics[width=1.0\textwidth,height=0.8\textheight]{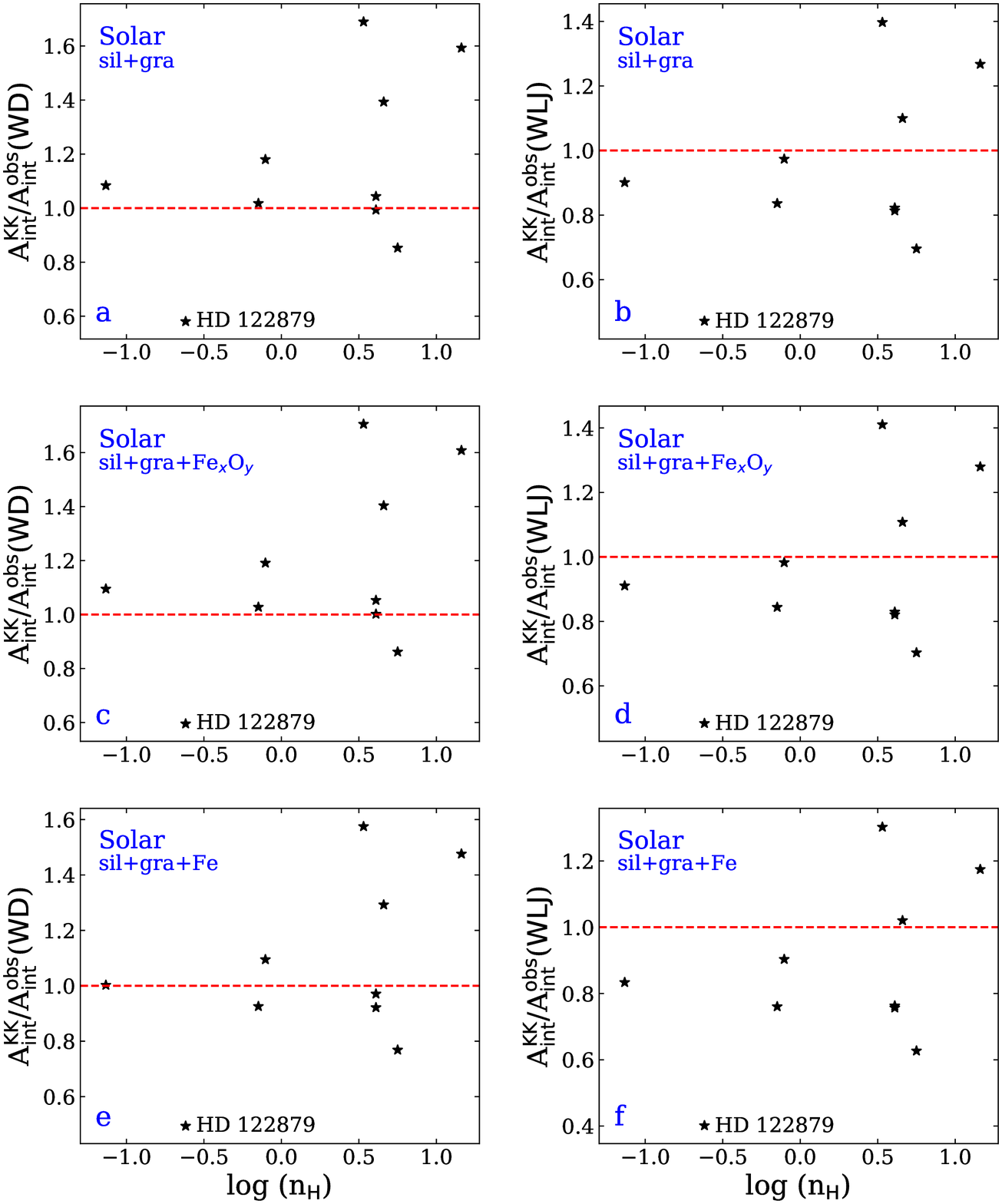}
	\caption{\footnotesize
	\label{fig:Solar}
	Same as Figure~\ref{fig:Bstar} 
        but with the solar abundances
        as the interstellar reference abundances.
	}
\end{figure*}

\begin{figure*}
	\centering	
	\includegraphics[width=1.0\textwidth,height=0.8\textheight]{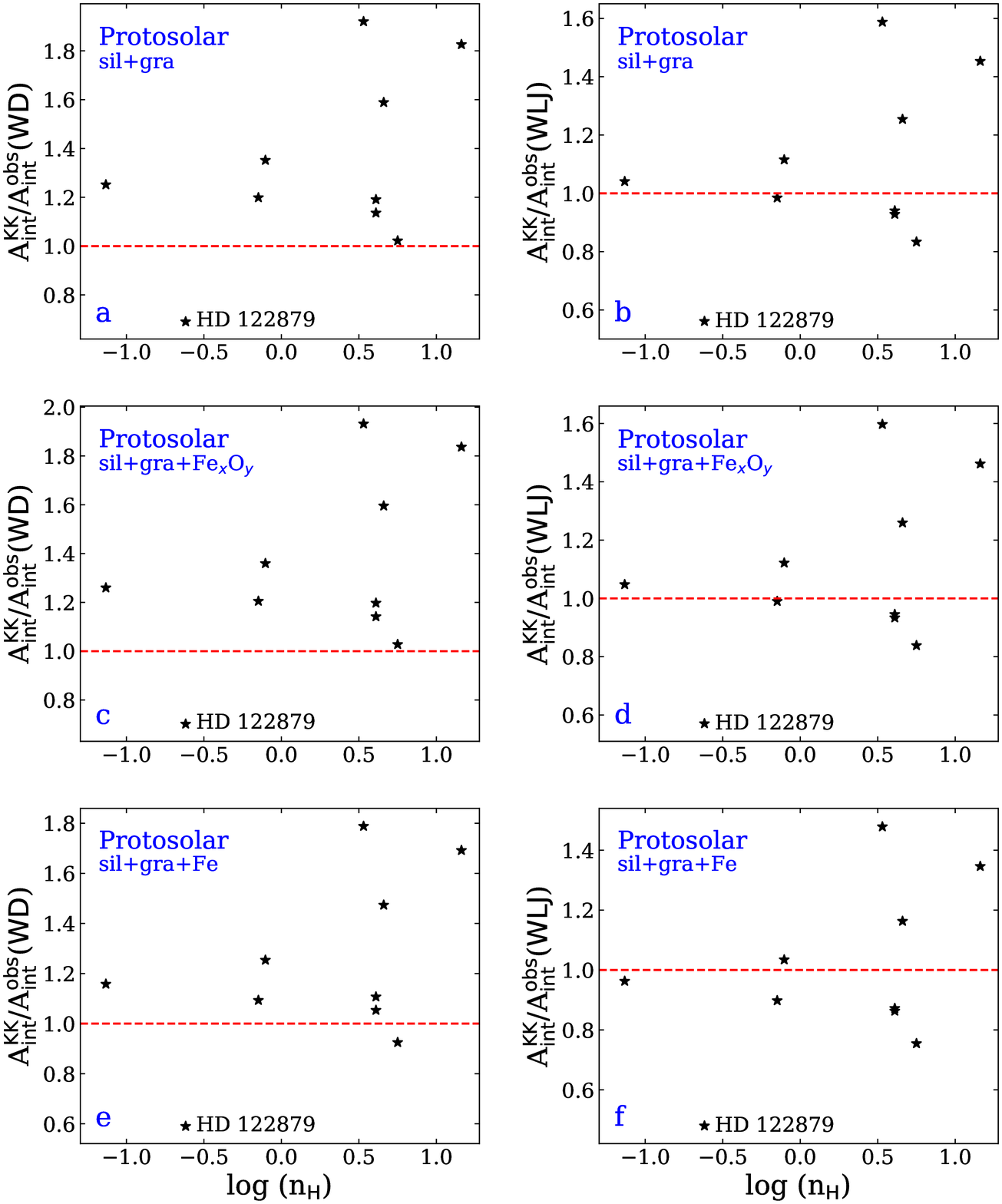}
	\caption{\footnotesize
	\label{fig:Protosolar}
        Same as Figure~\ref{fig:Bstar} 
        but with the protosolar abundances
        as the interstellar reference abundances.
	}
\end{figure*}

\begin{figure*}
	\centering	
	\includegraphics[width=1.0\textwidth,height=0.8\textheight]{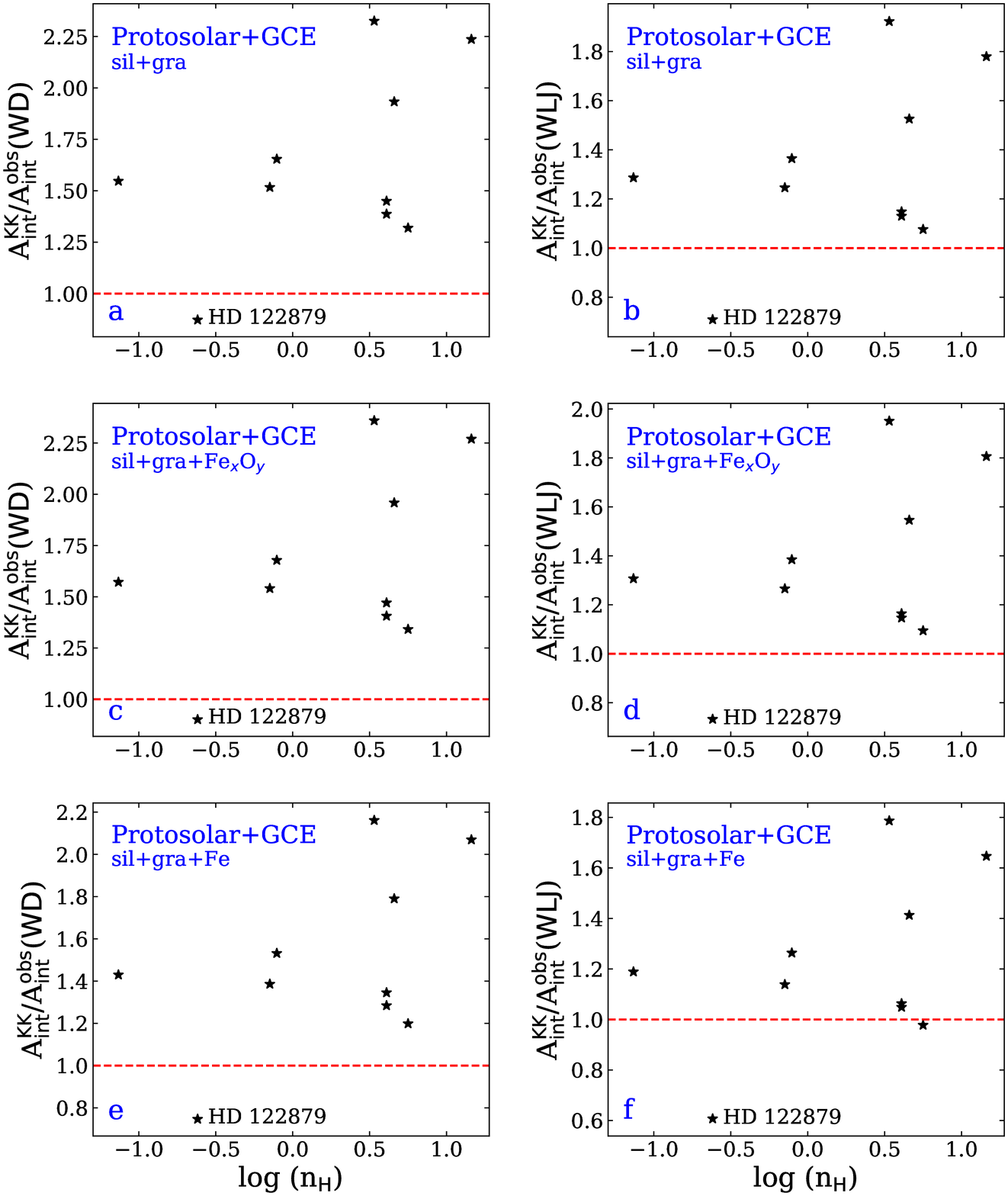}
	\caption{\footnotesize
	\label{fig:ProtosolarGCE}
        Same as Figure~\ref{fig:Bstar} 
        but with the GCE-augmented protosolar abundances
        as the interstellar reference abundances.
	}
\end{figure*}

We first consider the abundances of B stars. 
In Figure~\ref{fig:Bstar} we compare the
wavelength-integrated ``observed'' extinction
$\AintobsWD/\NH$ and $\AintobsWLJ/\NH$
with $\AintKK/\NH$, the KK-based extinction 
integrated over all wavelengths obtained by
assuming the interstellar abundances are that
of B stars. With $\AintKK/\NH$ exceeding 
$\AintobsWLJ/\NH$ for only one sightline 
(over 10 sightlines), it is apparent that 
the B-star abundances are not a viable 
representation of the interstellar abundances.
Even if we compare $\AintKK/\NH$ 
with $\AintobsWD/\NH$, we still find 
$\AintKK/\NH < \AintobsWD/\NH$
for the majority (7 over 10) of the sightlines.
This is true, irrespective of the exact form which
the Fe atoms missing from the gas phase take
(i.e., amorphous silicates, iron oxides or solid iron).
We have also considered the solar abundances
as the interstellar reference abundances.
As illustrated in Figure~\ref{fig:Solar}, 
we reach $\AintKK/\NH > \AintobsWLJ/\NH$ only for
a small fraction (3/10) of the sightlines.
Even if we assume the interstellar abundances to
be that of protosolar, the number of sightlines
with $\AintKK/\NH < \AintobsWLJ/\NH$ 
persists to be substantial.
As shown in Figure~\ref{fig:Protosolar}, 
half of the 10 sightlines still have 
$\AintKK/\NH < \AintobsWLJ/\NH$.
However, this changes when the GCE-augmented 
protosolar abundances are adopted as the interstellar
reference standard. 
As shown in Figure~\ref{fig:ProtosolarGCE},
we achieve $\AintKK/\NH > \AintobsWLJ/\NH$
for all the sightlines except HD\,122879.
The line of sight toward HD\,122879 has 
an unusually large gas-phase carbon abundance
of $\cgas\approx324\pm38\ppm$ (Parvathi et al.\ 2012). 
Such a high $\cgas$ abundance is difficult to 
reconcile with the observed extinction and other 
C-related interstellar spectral phenomena.\footnote{%
  If the interstellar C/H abundance is like
  the GCE-augmented protosolar C/H abundance
  of $\cism\approx339\pm39\ppm$,
  there will be only a small amount of C atoms 
  (i.e., $\cdust=\cism-\cgas\approx15\ppm$)
  left for making carbon dust and therefore
  it is not surprising that it leads to an unusually 
  small $\AintKK/\AintobsWLJ$ ratio.
  Such a low $\cdust$ abundance is troublesome
  since the ubiquitous and widespread
  ``unidentified'' IR emission (UIE) bands 
  at 3.3, 6.2, 6.2, 7.7, 8.6 and 11.3$\mum$ 
  {\it alone} require their carriers to lock up 
  $\simali$40--60$\ppm$ of C/H (see Li \& Draine 2001). 
  In addition, other interstellar spectral phenomena,
  e.g., the so-called extended red emission 
  (ERE, Witt \& Vijh 2004, Witt 2014), the 2175$\Angstrom$
  extinction bump (Draine 1989), and the 3.4$\mum$
  aliphatic C--H stretching absorption band 
  (Pendleton \& Allamandola 2002),
  also require an appreciable amount of 
  C/H to be tied up in their carriers. 
  Unless the line of sight toward HD\,122879
  is locally substantially enhanced in C, 
  it is difficult to account for the observed extinction 
  as well as the UIE bands, the 2175$\Angstrom$ 
  extinction bump, and the 3.4$\mum$ absorption band, 
  while in the mean time this sightline has 
  such a high $\cgas$ abundance.
  }
Therefore, we argue that the GCE-augmented 
protosolar abundances are a viable interstellar 
reference standard. 
%


So far, we have assumed the carbonaceous dust 
component to be graphite. However, other forms 
of carbonaceous solid materials 
(e.g.,  amorphous carbon)
have also been postulated to be a major 
interstellar dust component
(see Henning \& Salama 1998). 
Depending on their H contents, 
the mass densities of 
laboratory amorphous carbon materials 
range from $\simali$1.4$\g\cm^{-3}$ 
to $\simali$2.0$\g\cm^{-3}$ 
(see J\"{a}ger et al.\ 1998, 
Li \& Greenberg 2002).\footnote{%
   Jana et al.\ (2019) found that the mass
   density of amorphous carbon could
   range from $\simali$1.4$\g\cm^{-3}$ 
   to $\simali$3.5$\g\cm^{-3}$, with a typical
   value of $\simali$2.25$\g\cm^{-3}$,
   close to that of graphite. 
   }
On the other hand, those (H-rich) materials 
with a low mass density often have 
no or low DC conductivities (J\"{a}ger et al.\ 1998)
and therefore their static dielectric constants
$\varepsilon_0$ are much smaller
than that of graphite. This implies 
a smaller $F$ factor than that of graphite
for H-rich amorphous carbon.
Although a lower mass density leads to 
a larger dust volume (see Equation [\ref{eq:Vgra2H}]),
this will be compensated by a smaller $F$
so that the KK-based wavelength-integrated
$\AintKK$ will not be appreciably affected
(see Equation [\ref{eq:kk}]).
Even if we adopt the $F$ factor of graphite 
for amorphous carbon and assume a mass 
density of 1.8$\g\cm^{-3}$, the resulting 
$\AintKK$ would only increase by $\simali$10\%
and this would not affect our conclusion.

We have so far also assumed that the ISM consists 
of separate, distinct individual grain populations
(i.e., amorphous silicate, graphite, iron oxides, 
and iron). However, in the literature other dust 
structural forms have also been proposed, 
including silicate core-carbon mantle grains
(Jones et al.\ 1990, Li \& Greenberg 1997),
composite grains consisting of 
small silicates, amorphous carbon, 
and vacuum (Mathis 1996),
and composite ``astrodust'' consisting of
amorphous silicates, metal oxides, 
hydrocarbons, and vaccum (Draine \& Hensley 2020).
Nevertheless, whether the dust takes 
a core-mantle structure or a mixture
of separate, distinct individual grains would not
affect the dust volume as long as the amount
of dust material is fixed. 
On the other hand, compared to compact grains
of the same amount of material, a larger volume 
is expected for composite grains. 
However, composite grains will have 
a smaller $F$ due to the reduction of their static 
dielectric constant $\varepsilon_0$ (see Li 2005) 
and hence are not expected to result in a higher $\AintKK$.
Therefore, what exact structural form interstellar dust
may take does not affect our conclusion. 

\begin{figure*}
\centering
\includegraphics[width=1.0\textwidth,height=1.0\hsize]{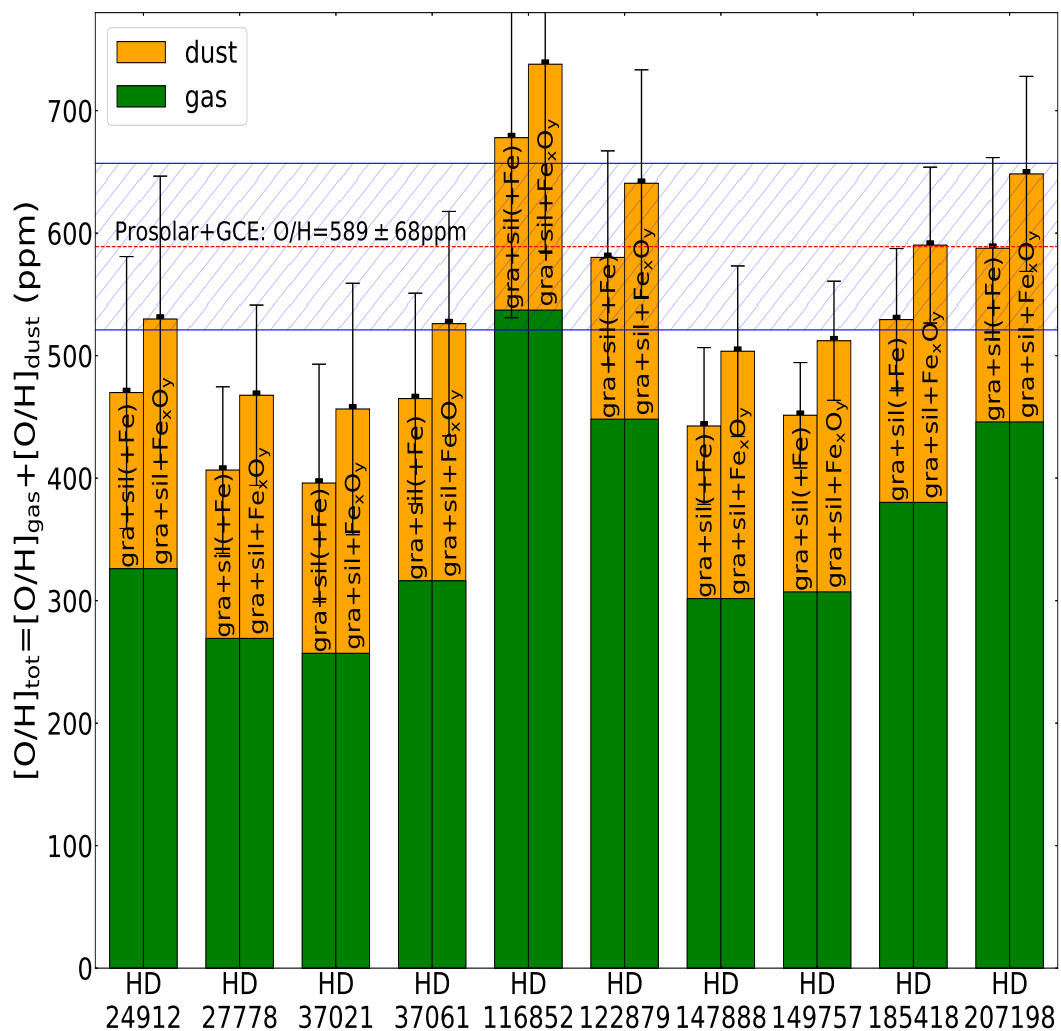}
\caption{\footnotesize
         \label{fig:O2H}
          Comparison of the GCE-augmented protosolar 
          O/H abundance (589$\pm$68$\ppm$; 
          blue horizontal shaded box)
          with $\otot$, the total amounts of O/H that could 
          be accommodated by gas (green vertical boxes) 
          and dust (orange vertical boxes)
          in each line of sight. 
          The (vertical) error bars are for $\otot$,
          resulting from the uncertainties in $\ogas$
          and the uncertainties in the GCE-augmented
          protoslar Si/H and Fe/H abundances. 
  	  }
\end{figure*}

As early as mid-1990s, it has been recognized that,
if the interstellar abundances are like that of B stars,
the amount of C atoms left for dust after subtracting
the gas-phase C abundance are insufficient to form 
the carbonaceous dust species required by dust models
(Snow \& Witt 1995, 1996).
This, known as the ``C crisis'', still holds
even if one assumes the interstellar C abundance
to be solar. For O, there is also a case of 
``O crisis'' or ``missing O''.
Unlike C which is insufficient, for O there is 
as much as $\simali$160$\ppm$ of O/H
unaccounted for in interstellar atoms, 
molecules and dust (Jenkins 2009, Whittet 2001a,b).
Wang et al.\ (2015b) suggested that $\mu$m-sized 
H$_2$O ice grains could accommodate the excess O/H
without exhibiting the 3.1$\mum$ absorption band of
H$_2$O ice and they could be present in the diffuse ISM
through rapid exchange of material
between dense clouds where they form
and diffuse clouds where they are destroyed 
by photosputtering.
Alternatively, H$_2$O ice could be trapped in silicates. 
Very recently, Potapov et al.\ (2020) found evidence for 
the trapping of H$_2$O ice in silicate grains based on
an analysis of the Spitzer/IRS spectra in combination with 
laboratory data.
We examine the depletion of O in the context
of the GCE-augmented protosolar abundances 
as the interstellar reference standard. 
If we assume that the ISM consists of graphite
and Fe-containing silicates 
(or Fe-lacking silicates plus iron grains), 
the total amount of O/H
which the ISM could accommodate is  
\begin{equation}\label{eq:O1}
\otot \approx \ogas + 3.5\times\left\{\siism-\sigas\right\}  ~~.
\end{equation}
Similarly, if the ISM consists of graphite
and Fe-lacking silicates
plus iron oxides, the ISM could accommodate 
a total amount of 
\begin{equation}\label{eq:O2}
\otot \approx \ogas + 3.5\times\left\{\siism-\sigas\right\} 
         + \frac{23}{18}\times\left\{\feism-\fegas\right\} ~~,
\end{equation}
where we assume equal amounts of 
FeO, Fe$_2$O$_3$ and Fe$_3$O$_4$.
Taking $\siism$ and $\feism$ to be that of 
the GCE-augmented protosolar abundances, 
for each line of sight, we calculate $\otot$
for different dust mixtures. 
In Figure~\ref{fig:O2H} we compare $\otot$ with
the GCE-augmented protosolar O/H abundance.
It appears that the majority of the sightlines have
no difficulty in accommodating the ``missing O'',
particularly if the Fe atoms are tied up in oxides.
Very recently, Psaradaki et al.\ (2020) compared 
the experimental X-ray spectra of the oxygen K edges 
of various silicate and oxide dust materials 
with the X-ray spectrum of Cygnus X-2, 
a bright low-mass X-ray binary,
observed by XMM-Newton. 
They derived a remarkably high gas-phase
abundance of $\ogas\approx610\pm60\ppm$
for the line of sight toward Cygnus X-2,
although an accurate derivation of 
the $\ogas$ abundance relies on 
an accurate knowldge of the atomic data 
of the oxygen edge spectral region.
Also, they determined the solid-phase
O/H abundance to be $\odust\approx45\pm7\ppm$,
which is smaller than that could be accommodated 
by silicate dust alone by a factor of $\simali$3.
Nevertheless, the total O/H abundance falls in 
the high side of the GCE-augmented protosolar
O/H abundance.

\begin{figure*}
\centering
\includegraphics[width=1.0\textwidth,height=1.0\hsize]{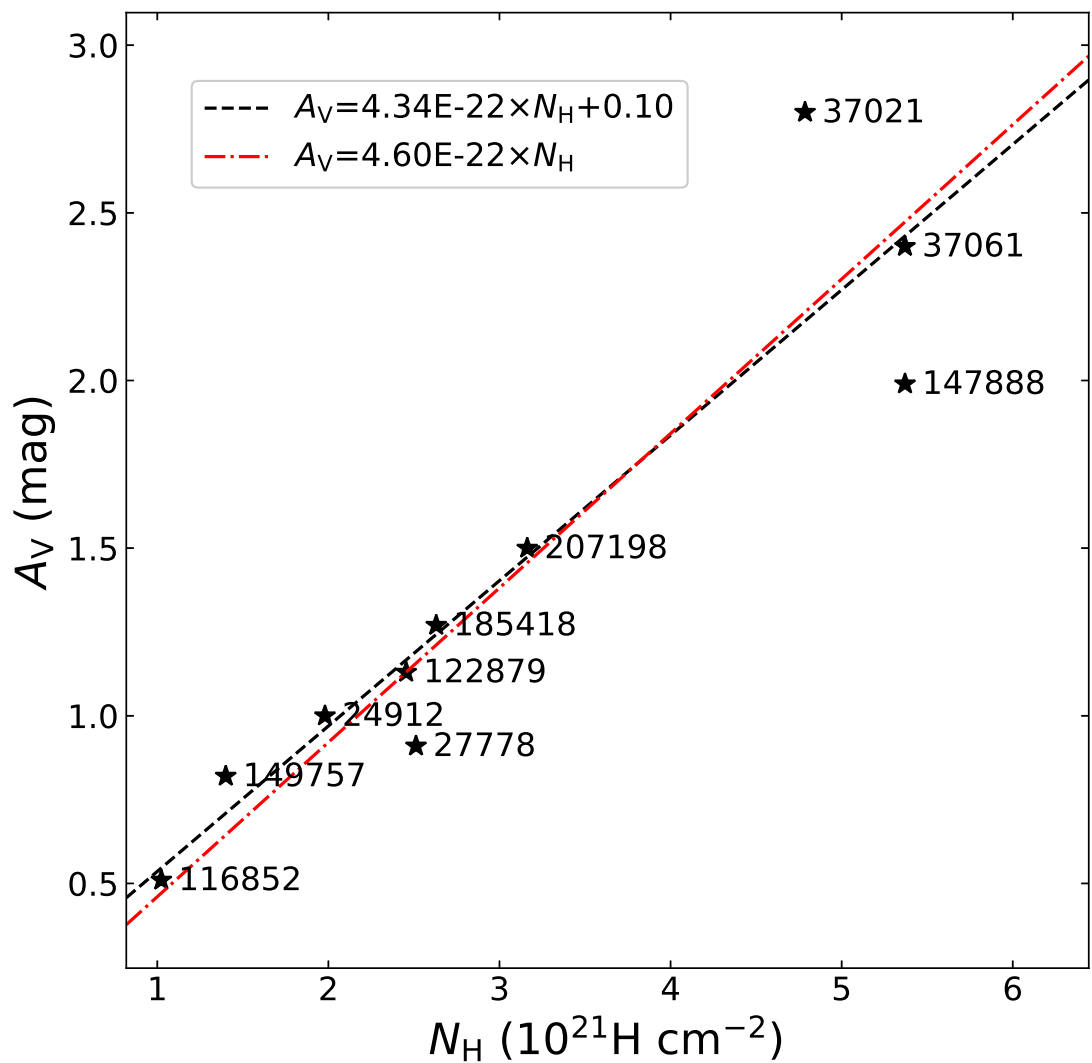}
\caption{\footnotesize
             \label{fig:AV2NH}
             Visual extinction ($\AV$) against
             the hydrogen column densities ($\NH$)
             for the ``gold'' sample of 10 sightlines
             for which both the extinction parameters
             and the gas-phase abundances of C, O,
             Si, Mg and Fe have been observationally determined.
             Broken lines fit a linear relationship to the data. 
             The goodness of fit is measured by
              $\chi^2/\dof$, the chi-square per degree of freedom
              ($\chi^2/\dof\approx0.35$ for black dashed line
              and $\chi^2/\dof\approx0.37$ for red dot-dashed line).
       	       }
\end{figure*}

It is widely believed that in the ISM dust and gas 
are well mixed, as evidenced by the tight empirical 
correlation between reddening $E(B-V)$
and hydrogen column density $\NH$.
Bohlin et al.\ (1978) derived 
the hydrogen-to-reddening ratio to be 
$\NH/E(B-V)\approx5.8\times10^{21}\,\rmH\cm^{-2}\magni^{-1}$
for a sample of 100 stars with 
$E(B-V)$ up to $\simali$0.5$\magni$,
based on the UV absorption spectra
of HI and H$_2$ observed by the {\it Copernicus} satellite.
With $\RV\approx3.1$ for the Galactic diffuse ISM,
this corresponds to 
$\AV/\NH\approx5.3\times10^{-22}\magni\cm^{2}\,\rmH^{-1}$,
a ratio long taken to be representative of the ISM
in the solar neighbourhood.
However, appreciably lower extinction-to-hydrogen ratios
have been reported (see Hensley \& Draine [2020b]
and references therein), e.g., Lenz et al.\ (2017) recently
derived $\AV/\NH\approx3.5\times10^{-22}\magni\cm^{2}\,\rmH^{-1}$
for diffuse, low-column-density regions 
with $\NHI < 4\times 10^{20}\,\rmH\cm^{-2}$.
We have also examined the $\AV$--$\NH$ relation
for the 10 sightlines of our ``gold'' sample 
compiled in this work and determined
$\AV/\NH\approx4.6\times10^{-22}\magni\cm^{2}\,\rmH^{-1}$
(see Figure~\ref{fig:AV2NH}).
This extinction-to-hydrogen ratio is intermediate 
between that of Bohlin et al.\ (1978) 
and that of Lenz et al.\ (2017) 
and close to that of Zhu et al.\ (2017)
who derived 
$\AV/\NH\approx4.8\times10^{-22}\magni\cm^{2}\,\rmH^{-1}$
from X-ray observations
of a large sample of Galactic sightlines
toward supernova remnants, 
planetary nebulae, and X-ray binaries.

\begin{figure*}
\vspace{-0.5cm}
\begin{minipage}[t]{0.5\textwidth}
\resizebox{9.0cm}{7.0cm}{\includegraphics[clip]{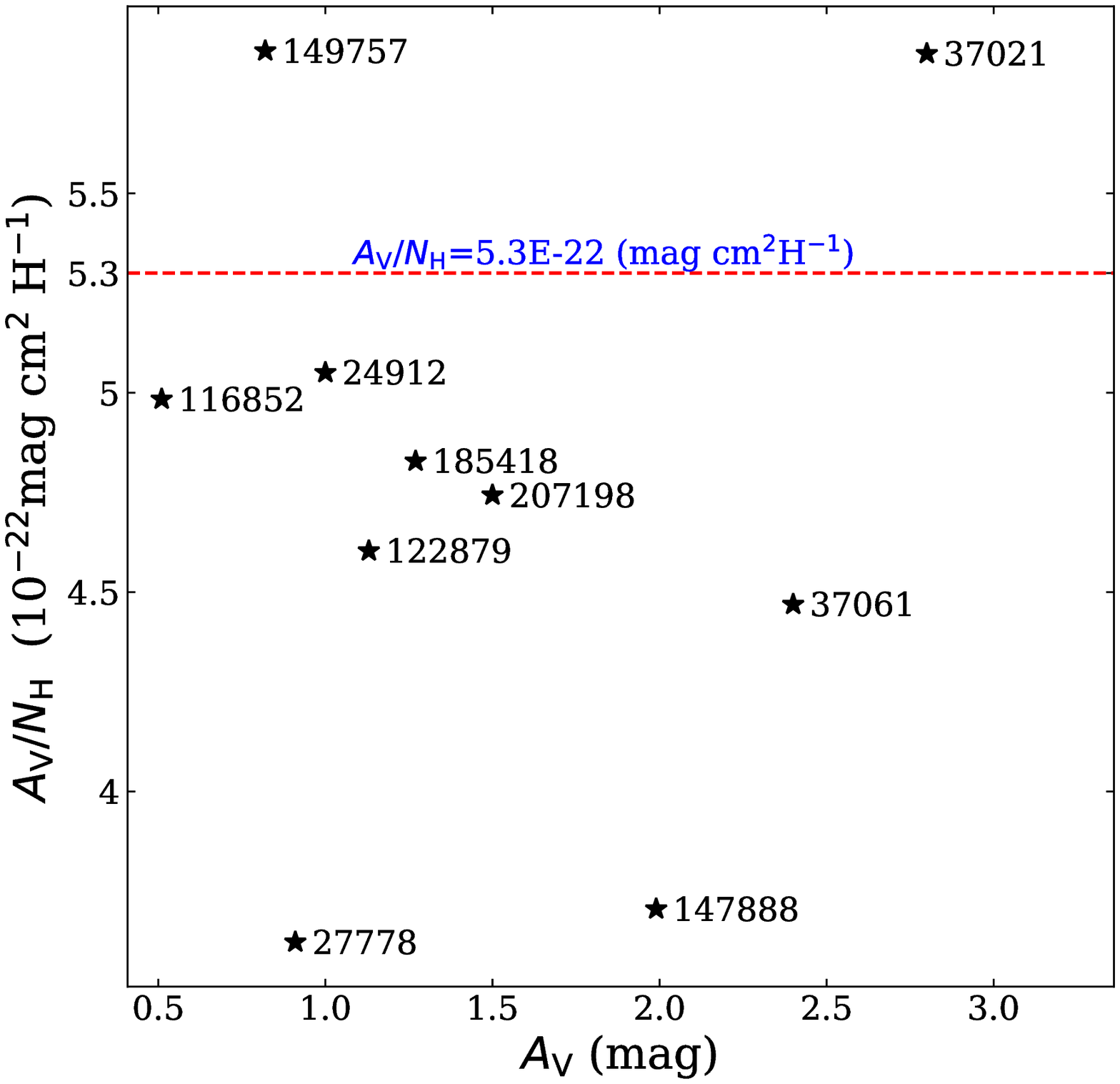}}\vspace{-0.3cm}
\resizebox{9.0cm}{7.0cm}{\includegraphics[clip]{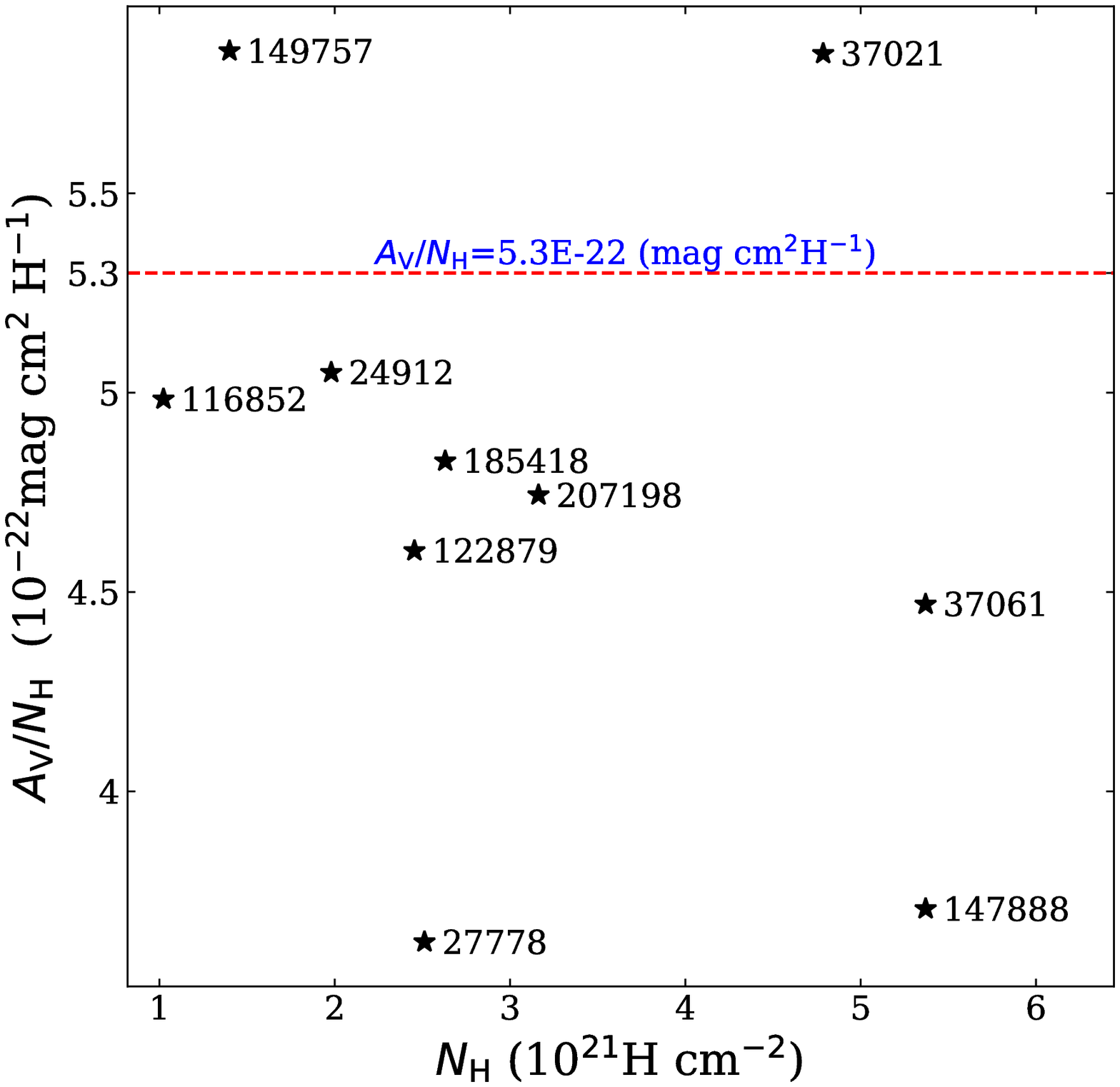}}\vspace{-0.3cm}
\end{minipage}\hspace{-1.2cm}
\begin{minipage}[t]{0.5\textwidth}
\resizebox{9.0cm}{7.0cm}{\includegraphics[clip]{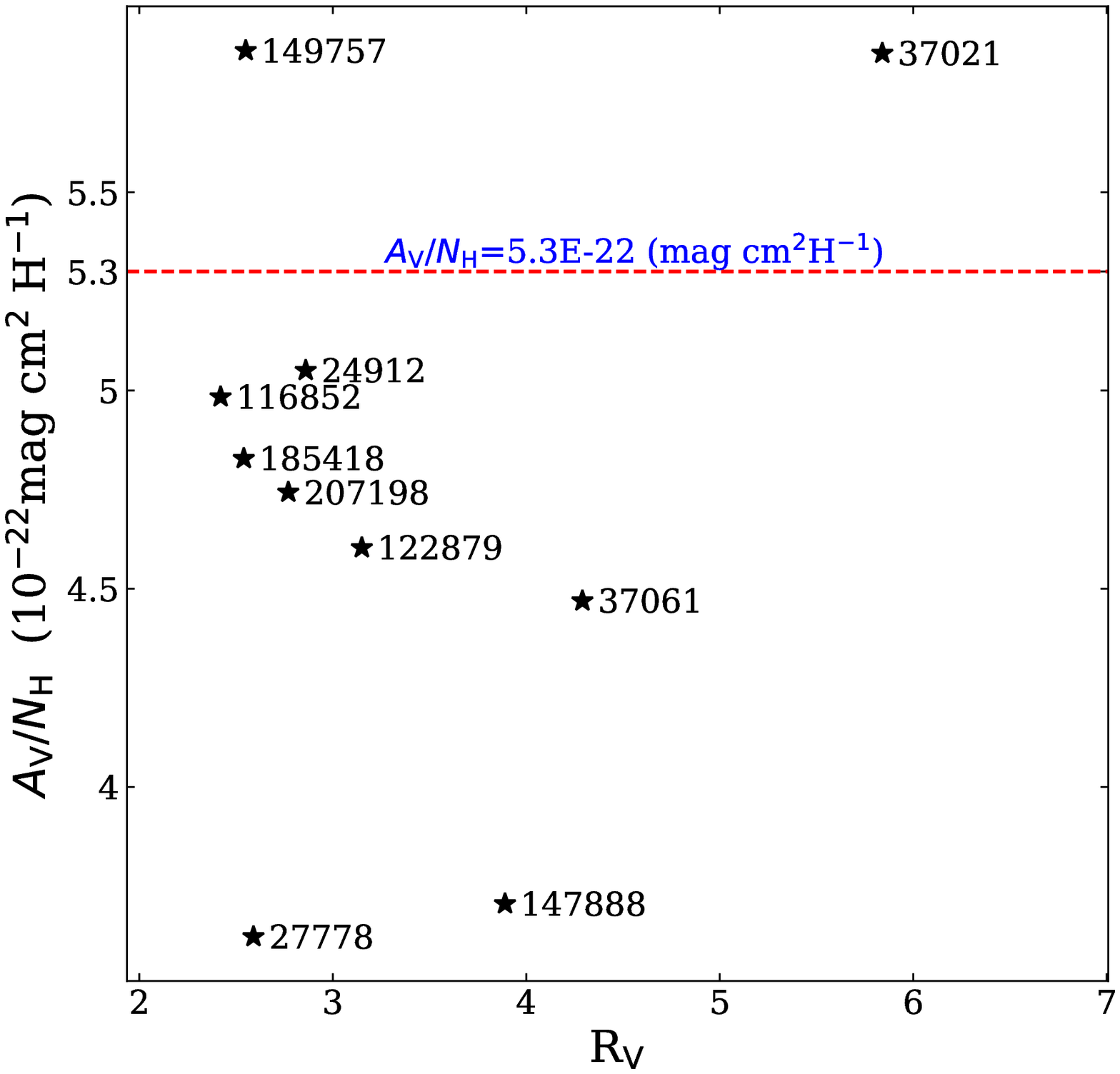}}\vspace{-0.3cm}
\resizebox{9.0cm}{7.0cm}{\includegraphics[clip]{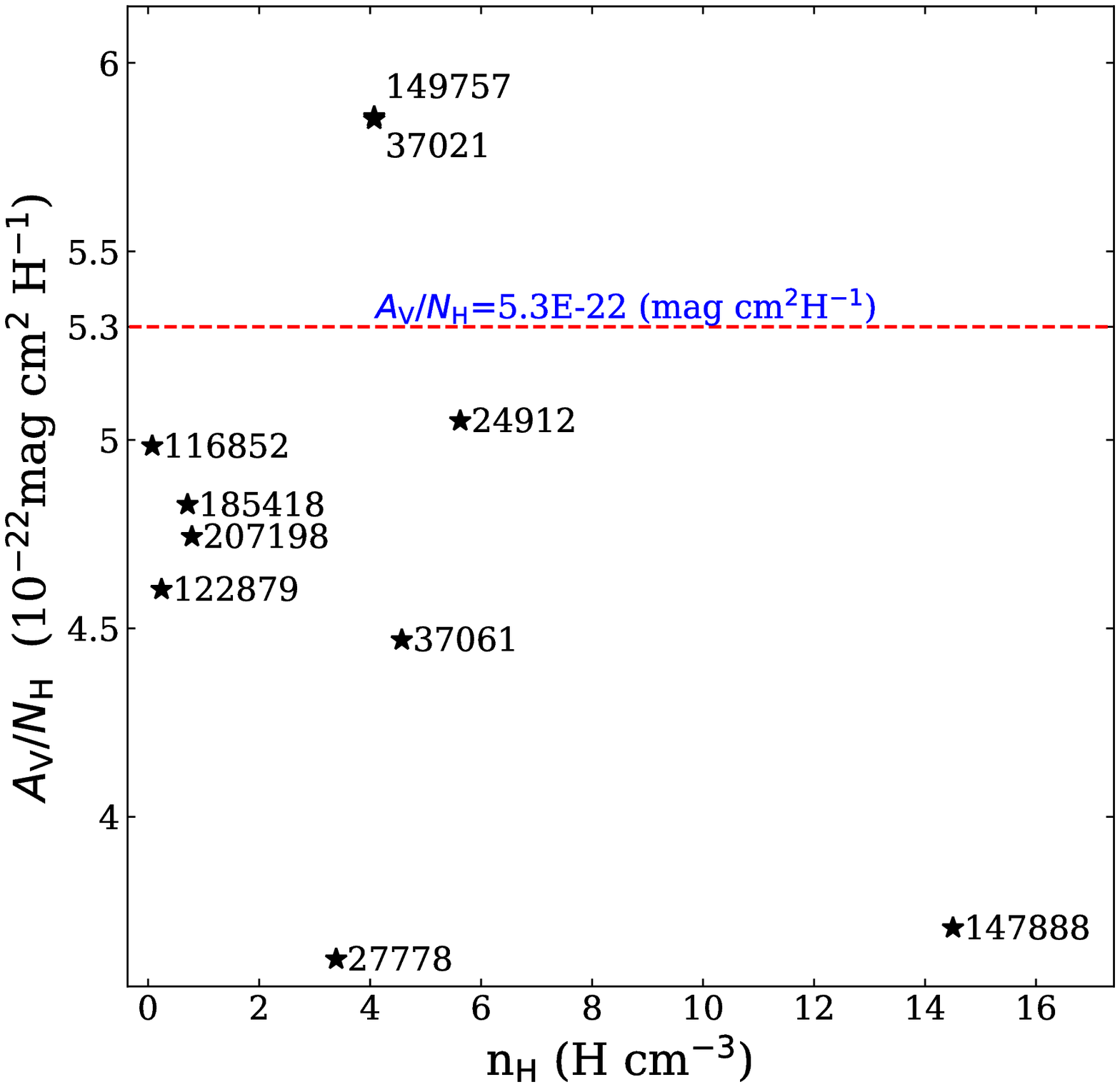}}\vspace{-0.3cm}
\end{minipage}\hspace{-1.2cm}
\begin{minipage}[t]{0.5\textwidth}
\resizebox{9.0cm}{7.0cm}{\includegraphics[clip]{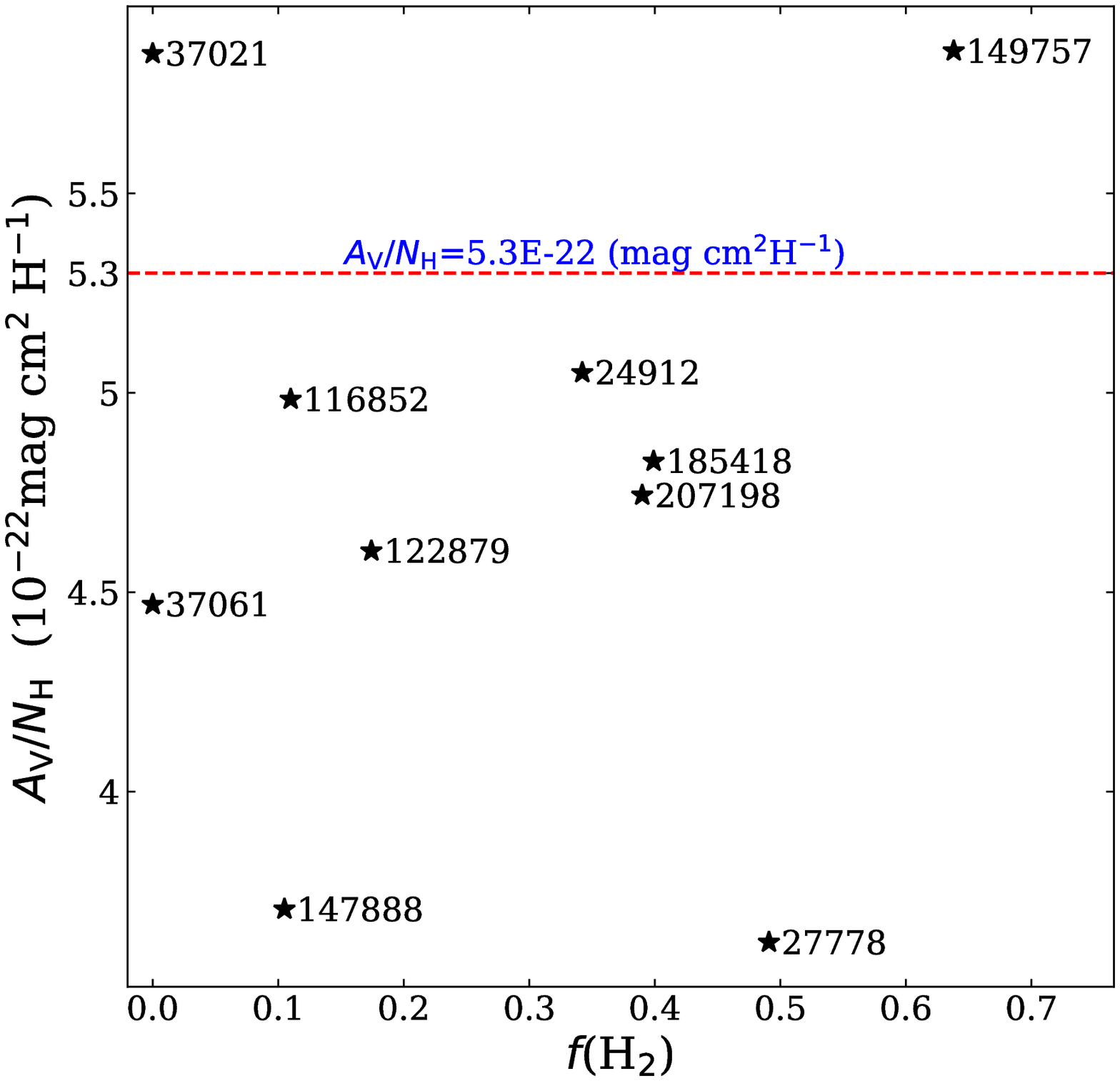}}\vspace{-0.3cm}
\end{minipage}
\caption{\footnotesize
         \label{fig:AVNHvsfH2}
         Relations of the ratio of visual extinction 
         to hydrogen column density ($\AV/\NH$)
         with the visual extinction ($\AV$; panel a),
         the total-to-selective extinction ratio ($\RV$; panel b),
         the hydrogen column density ($\NH$; panel c),
         the hydrogen volume density ($\nH$; panel d), and
         the molecular hydrogen fraction 
         [$f(\Htwo)\equiv 2\NHH/\left\{\NHI 
          + 2\NHH\right\}$, panel e].
         The dashed horizontal line shows
         the canonical ratio of 
         $\AV/\NH=5.3\times10^{-23}\magni\cm^2\HH^{-1}$. 
         }
\end{figure*}

We have also explored the relationships 
between $\AV/\NH$ 
and the physical and chemical conditions 
of the interstellar environments, 
including $\AV$, $\RV$,
$\NH$, $\langle n_{\rm H}\rangle$,  
and $f(\Htwo)\equiv 
2\NHH/\left\{\NHI + 2\NHH\right\}$,
the fraction of molecular hydrogen
in the line of sight 
(see Figure~\ref{fig:AVNHvsfH2}).
For the 10 lines of sight considered here,
$\AV/\NH$ does not appear to show any
appreciable correlations with these parameters. 
This seems to contradict the conventional belief
that, due to grain aggregation (e.g., see Jura 1980),
$\AV/\NH$ decreases toward denser regions
which are often characterized by larger values 
of $\RV$, $\langle n_{\rm H}\rangle$, and $f(\Htwo)$.
%
Kim \& Martin (1996) explored the variation of
$\AV/\NH$ with $\RV$ for several dozen
sightlines spanning $2.7<\RV<5.6$. 
Despite a large scatter,
they found a somewhat increase of 
$\AV/\NH$ with $\RV$ for $\RV<4.4$,
and above this value, $\AV/\NH$ tends 
to be smaller. 
With $\RV\approx2.1$, 
the high Galactic latitude cloud 
toward HD\,210121 has 
$\AV/\NH\approx4.2\times10^{-22}\magni\cm^{2}\,\rmH^{-1}$
(see Li \& Greenberg 1998), about 20\% lower
than that of the canonical value of
$\AV/\NH\approx5.3\times10^{-22}\magni\cm^{2}\,\rmH^{-1}$
(Bohlin et al.\ 1978).\footnote{%
  The extinction curves of the interstellar lines 
  of sight toward Type Ia supernovae
  in external galaxies are often very steep 
  and characterized by $\RV<2$
  (see Howell 2011).
  However, there is often very little information
  on $\AV/\NH$.
  }

\section{Summary}\label{sec:summary}
We have compiled a ``gold'' sample of 10 lines of sight
for which the extinction parameters and the gas-phase
abundances of the dust-forming elements C, O, Si, Mg 
and Fe have been observationally determined. 
We have applied the KK relation to this sample 
to examine the viability of 
(i) the abundances of B stars, 
(ii) the solar, 
(iii) protosolar and 
(iv) GCE-augmented solar abundances 
as the interstellar reference abundances. 
Except that we assume the ISM is made of
either 
(i) graphite and Fe-containing amorphous silicates,
or 
(ii) graphite and Fe-lacking amorphous silicates
plus iron oxides,
or 
(iii) graphite and Fe-lacking amorphous silicates 
plus iron, 
our approach is model-independent 
in the sense that we do not need to specify
the dust size distritbution and do not need 
to reproduce the observed extinction curves. 
Our principal results are as follows:
\begin{enumerate}
\item For each of the (10) lines of sight, 
          each of the (four) assumed sets
          of interstellar reference abundances
          and each of the (three) assumed 
          different dust mixtures, 
          we have calculated the dust volumes and
          the KK-based wavelength-integrated extinction. 
          We have also calculated the observation-based
          wavelength-integrated extinction for each sightline.
          We have found that only the GCE-augmented 
          protosolar abundances could meet the KK criterion 
          that the former must exceed the latter. 
\item Although we have assumed the interstellar carbonaceous
          dust component to be graphite, the exact composition of
          this component (e.g., amorphous carbon vs. graphite) 
          is not critical. 
          Also, although we have assumed interstellar dust to be 
          a mixture of separate, distinct individual dust species, 
          the exact dust structural form 
          (e.g., core-mantle or composite) 
          does not affect our conclusion.
\item For this sample we have investigated 
          the ``missing O'' problem
          (i.e., in the diffuse ISM a substantial fraction of
          the O/H remains unaccounted for in interstellar 
          atoms, molecules and dust) and found that for 
          the majority of the lines of sight the ISM does not 
          seem to have difficulty in accommodating the O atoms.
\item For this sample we have derived 
          the extinction-to-hydrogen gas ratio to be
          $\AV/\NH\approx4.6\times10^{-22}\magni\cm^{2}\,\rmH^{-1}$,
          $\simali$13\% lower than the canonical value of
          $\AV/\NH\approx5.3\times10^{-22}\magni\cm^{2}\,\rmH^{-1}$.
          Also, $\AV/\NH$ does not systematically decrease
          toward denser regions as indicated by larger values
          of $\RV$, hydrogen volume density 
          $\langle n_{\rm H}\rangle$,  
          and molecular fraction of hydrogen $f(\Htwo)$,
          contrary to the conventional wisdom of more reduced
          $\AV/\NH$ in denser regions due to grain coagulation.         
\end{enumerate}

\acknowledgments
WBZ and GZ are supported in part 
by the National Natural Science Foundation of China 
under grants No.\,11988101 and No.\,11890694 
as well as the National Key R\&D Program of China 
(No.\,2019YFA0405502).
We thank A.~Mishra for helpful discussions
and the anonymous referee for 
his/her constructive suggestions.

\vspace{-3mm}

\clearpage

\begin{table}[h!]
\begin{center}
\footnotesize
\caption{\label{tab:abund}
                Solar and stellar abundances 
                as plausible interstellar reference
                abundances for the dust-forming elements 
                (relative to 10$^6$ H atoms).}
\begin{tabular}{lccccr}
\hline \hline
Element  & B stars$^{a}$
         & Solar$^{b}$ 
         & Solar$^{c}$
         & Protosolar$^{d}$ 
         & GCE-augmented protosolar$^{e}$\\ 
\hline
C     & $209\pm15$ 
   & $363\pm33$  
   & $269\pm31$ 
   & $288\pm27$    
   & $339\pm39$\\
O     & $575\pm40$ 
   & $851\pm69$ 
   & $490\pm57$  
   & $575\pm66$ 
   & $589\pm68$\\    
Mg    & $36.3\pm4.2$ 
   & $38.0\pm4.4$
   & $39.8\pm3.7$ 
   & $41.7\pm1.9$ 
   & $47.9\pm4.4$\\ 
Si    & $31.6\pm1.5$
   & $35.5\pm4.1$
   & $32.4\pm2.2$
   & $40.7\pm1.9$
   & $42.7\pm4.0$\\
Fe    & $27.5\pm2.5$ 
   & $46.8\pm3.2$  
   & $31.6\pm2.9$
   & $34.7\pm2.4$
   & $47.9\pm4.4$\\
\hline
\end{tabular}
\\
(a) Przybilla et al.\ (2008); 
(b) Anders \& Grevesse (1989); 
(c) Asplund et al.\ (2009); 
(d) Lodders (2003); 
(e) Protosolar abundances augmented by GCE 
    (Asplund et al.\ 2009, Chiappini et al.\ 2003). 
\end{center}
\end{table}

\thispagestyle{empty}
\setlength{\voffset}{25mm}
\begin{deluxetable}{lccccccccr}
\rotate 
\tablecolumns{10}
\tabletypesize{\tiny}
\tablewidth{0truein}
\center
\tablecaption{Gas Column Densities and Abundances 
                       of the Ten Interstellar Lines of Sight 
                       in the ``Gold'' Sample
                      \label{tab:GasAbund}
                      }
\tablehead{
\colhead{Star}&
\colhead{$\NH$}&
\colhead{$\NHI$}&
\colhead{$\NHH$}&
\colhead{$\cgas$}&
\colhead{$\ogas$}&
\colhead{$\sgas$}&
\colhead{$\mggas$}&
\colhead{$\sigas$}&
\colhead{$\fegas$}
\\
\colhead{}&
\colhead{($10^{21}\cm^{-2}$)}&
\colhead{($10^{21}\cm^{-2}$)}&
\colhead{($10^{21}\cm^{-2}$)}&
\colhead{(ppm)}&
\colhead{(ppm)}&
\colhead{(ppm)}&
\colhead{(ppm)}&
\colhead{(ppm)}&
\colhead{(ppm)}
}
\startdata

  HD 24912 & 1.98$\pm$0.54 (2) & 1.20$\pm$0.18 (3) & 0.34$\pm$0.07 (3) & 163.1$\pm$86.5 (16) & 326.1$\pm$97.5 (3) & 1.75$\pm$0.48 (4) & 1.99$\pm$0.17 (2) & 1.61$\pm$0.44 (2) & 0.92$\pm$0.25 (13)\\
  HD 27778 & 2.51$\pm$0.44 (7) & 0.89 (5) & 0.62 (10) & 79.2$\pm$27.6 (14) & 269.2$\pm$ 53.8 (3) & --& 1.05$\pm$0.22 (7) & 3.43$\pm$ 0.60 (9) & 0.10$\pm$0.018 (9)\\
  HD 37021 & 4.79$\pm$1.50 (3) & 4.79$\pm$1.50 (3) & -- & 90.9$\pm$37.9 (14) & 257.0$\pm$82.6 (3) & --& 1.78$\pm$0.57 (7) & 2.99$\pm$1.15 (9) & 0.58$\pm$0.19 (9)\\
  HD 37061 & 5.37$\pm$1.20  (3) & 5.37$\pm$1.20 (3) &--& 98.1$\pm$35.5 (14) & 316.2$\pm$72.2 (3) & 0.38$\pm$17.9 (4) & 1.12$\pm$0.26 (7) & 0.20$\pm$0.17 (4) & 0.06$\pm$0.08 (4)\\  
  HD 116852 & 1.02$\pm$0.09 (1) & 0.91$\pm$0.08 (1) & 0.06$^{+0.012}_{-0.009}$ (1) & 117.3$\pm$103.2 (15) & 537.5$\pm$133.2 (1) & 3.89$\pm$27.4 (4) & 7.76$\pm$0.74 (1) & 2.57$\pm$0.39 (4) & 0.98$\pm$0.35 (4)\\
  HD 122879 & 2.45$^{+0.28}_{-0.23}$  (1) & 2.04$^{+0.28}_{-0.24}$  (1) &0.21$^{+0.04}_{-0.03}$ (1) & 324.3$\pm$38.2  (15) & 448.1$\pm$72.8 (1) &--& 6.44$\pm$0.76  (1) & 4.97$\pm$0.61  (2) & 0.53$\pm$0.10 (6)\\
  HD 147888 & 5.37$^{+0.87}_{-2.10}$ (1) & 4.79$^{+0.88} _{-2.10}$  (1) & 0.28$^{+0.03}_{-0.05}$  (1) & 105.4$\pm$22.6 (14) & 301.7$\pm$50.6 (1) & 0.49$\pm$0.33 (4) & 1.86$\pm$0.33 (1) & 2.44$\pm$0.60 (9) & 0.15$\pm$0.08 (6)\\        
  HD 149757 & 1.40$\pm$0.03 (2) & 0.51$\pm$0.02 (3) & 0.45$\pm$0.06 (3) & 100.9$\pm$48.6 (11) & 307.1$\pm$29.1 (8) & 2.37$\pm$4.14 (4) & 1.86$\pm$0.16 (2) & 1.50$\pm$0.04 (2) & 0.32$\pm$0.09 (4)\\
  HD 185418 & 2.63$^{+0.24}_{-0.18}$ (1) & 1.55$^{+0.18}_{-0.14} $ (1) & 0.53$^{+0.09}_{-0.07}$ (1) & 167.7$\pm$22.8 (15) & 380.2$\pm$43.8  (1) & 0.87$\pm$ 0.08 (12) & 4.07$\pm$0.42 (1) & 0.060$\pm$0.01 (12) & 0.32$\pm$0.09  (12)\\
  HD 207198 & 3.16$^{+0.36}_{-0.44} $ (1) & 1.91$^{+0.31}_{-0.40}$ (1) & 0.62$\pm$0.07 (1) & 102.1$\pm$25.1 (14) & 445.9$\pm$59.9 (1) & 1.20$\pm$20.2 (4) & 3.64$\pm$0.43 (1) & 2.19$\pm$1.07 (9) & 0.43$\pm$0.08 (9)\\

\enddata 
\\
(1) Jenkins (2019); 
(2) Gnaci\'nski et al.\ (2006); 
(3) Cartledge et al.\ (2004); 
(4) van Steenberg et al.\ (1988); 
(5) Welty et al.\ (2010); 
(6) Jensen et al.\ (2007a); 
(7) Jensen et al.\ (2007b);
(8) Knauth et al.\ (2006);
(9) Miller et al.\ (2007);
(10) Sheffer et al.\ (2007);
(11) Sofia et al.\ (1994);
(12) Sonnentrucker et al.\ (2003);
(13) Jenkins et al.\ (1986);
(14) Sofia et al.\ (2011);
(15) Parvathi et al.\ (2012); 
(16) Sofia et al.\ (2004).
\end{deluxetable}

\thispagestyle{empty}
\setlength{\voffset}{25mm}
\begin{deluxetable}{lcccccccccccccr}
\rotate 
\tablecolumns{15}
\tabletypesize{\tiny}
\tablewidth{0truein}
\center
\tablecaption{\footnotesize
              \label{tab:extpara}
              Extinction Parameters
              of the Ten Interstellar Lines of Sight 
              in the Gold Sample
              }
\tablehead{
\colhead{Star}&
\colhead{$\AUU$$^b$}&
\colhead{$\AB$$^b$}&
\colhead{$\AV$$^a$}&
\colhead{$\AJ$$^b$}&
\colhead{$\AH$$^b$}&
\colhead{$\AK$$^b$}&

\colhead{$E(B-V)$$^a$}&
\colhead{$\RV$$^a$}&
\colhead{$c_{1}^{\prime}$$^a$}&
\colhead{$c_{2}^{\prime}$$^a$}&
\colhead{$c_{3}^{\prime}$$^a$}&
\colhead{$c_{4}^{\prime}$$^a$}&
\colhead{$x_{0}$$^a$}&
\colhead{$\gamma$$^a$}
\\
\colhead{}&
\colhead{(mag)}&
\colhead{(mag)}&
\colhead{(mag)}&
\colhead{(mag)}&
\colhead{(mag)}&
\colhead{(mag)}&

\colhead{(mag)}&
\colhead{}&
\colhead{}&
\colhead{}&
\colhead{}&
\colhead{}&
\colhead{($\um^{-1}$)}&
\colhead{($\um^{-1}$)}
}
\startdata

  HD 24912 &--&--& 1.00$\pm$0.21 &--&--&--& 0.35$\pm$0.04 & 2.86$\pm$0.51 & 1.187$\pm$0.728 & 0.270$\pm$0.054 & 0.943$\pm$0.219 & 0.050$\pm$0.024 & 4.541$\pm$0.016 & 0.846$\pm$0.028\\
  HD 27778$^1$ &1.58&1.25& 0.91$\pm$0.14 &0.19&0.11&0.07& 0.35$\pm$0.04 & 2.59$\pm$0.24 & 1.421$\pm$0.252 & 0.232$\pm$0.038 & 0.878$\pm$0.180 & 0.386$\pm$0.062 & 4.603$\pm$0.012 & 0.974$\pm$0.032\\
  HD 37021 &--&--& 2.80$\pm$0.17 &--&--&--& 0.48$\pm$0.02 & 5.84$\pm$0.26 & 1.063$\pm$1.047 & 0.020$\pm$0.008 & 0.235$\pm$0.043 & 0.007$\pm$0.007 & 4.584$\pm$0.047 & 1.081$\pm$0.036\\
  HD 37061$^1$ &3.33&2.93& 2.40$\pm$0.21 &0.72&0.45&0.3& 0.56$\pm$0.04 & 4.29$\pm$0.21 & 1.544$\pm$0.145 & 0.000$\pm$0.100 & 0.310$\pm$0.042 & 0.050$\pm$0.012 & 4.574$\pm$0.014 & 0.901$\pm$0.029\\
  HD 116852 &--&--& 0.51$\pm$0.12 &--&--&--& 0.21$\pm$0.04 & 2.42$\pm$0.37 & 0.518$\pm$0.249 & 0.376$\pm$0.103 & 0.633$\pm$0.173 & 0.010$\pm$0.015 & 4.548$\pm$0.041 & 0.782$\pm$0.069\\
  HD 122879 $^2$&1.69&1.45& 1.13$\pm$0.20 &--&--&--& 0.36$\pm$0.05 & 3.15$\pm$0.30 & 1.321$\pm$0.299 & 0.233$\pm$0.040 & 1.243$\pm$0.230 & 0.190$\pm$0.039 & 4.581$\pm$0.004 & 0.831$\pm$0.021\\
  HD 147888 $^1$&2.85&2.48& 1.99$\pm$0.18 &0.6&0.35&0.22& 0.51$\pm$0.04 & 3.89$\pm$0.20 & 1.471$\pm$0.267 & 0.037$\pm$ 0.012 & 0.665 $\pm$0.100 & 0.087$\pm$0.022 & 4.587$\pm$0.013 & 0.879$\pm$0.029\\
  HD 149757 $^1$&1.26&1.09& 0.82$\pm$0.13 &0.27&0.14&0.1& 0.32$\pm$0.04 & 2.55$\pm$0.24 & 1.002$\pm$0.100 & 0.286$\pm$0.037 & 1.872$\pm$0.313 & 0.215$\pm$0.052 & 4.552$\pm$0.010 & 1.186$\pm$0.042\\
  HD 185418 $^1$&2.07&1.78& 1.27$\pm$0.14 &0.21&0.11&0.07& 0.50$\pm$0.04 & 2.54$\pm$0.20 & 1.817$\pm$0.265 & 0.100$\pm$0.018 & 1.156$\pm$0.170 & 0.158$\pm$0.029 & 4.604$\pm$0.005 & 0.819$\pm$0.024\\
  HD 207198 $^2$&2.43&1.96& 1.50$\pm$0.29 &--&--&--& 0.54$\pm$0.08 & 2.77$\pm$0.35 & 0.811$\pm$0.259 & 0.344$\pm$0.050 & 0.976$\pm$0.169 & 0.277$\pm$0.045 & 4.596$\pm$0.006 & 0.883$\pm$0.024\\

\enddata 
\\
\flushleft{$^{a}$ Data taken from Valencic et al.\ (2004). \\
$^{b}$ The U, B, J, H, K broadband photometric 
extinction data ($\AUU$, $\AB$, $\AJ$, $\AH$ and $\AK$) 
are taken from Fitzpatrick \& Massa (2007) for those lines
of sight marked by ``1'' and from Gordon et al.\ (2009)
for those marked by ``2''.  
}
\end{deluxetable}

\thispagestyle{empty}
\setlength{\voffset}{25mm}
\begin{deluxetable}{lccccccccccccccccc}
	\rotate 
	\tablecolumns{18}
	\tabletypesize{\tiny}
	\tablewidth{0truein}
	\center
	\tablecaption{\footnotesize
		\label{tab:Bstar}
                Wavelength-Integrated Extinction
                Obtained from the Observed Extinction Curves
                [$\AintobsWD$ and $\AintobsWLJ$]
                and from the Dust Volumes Based on
                the KK Relation [$\AintKK$].
                The Dust Volumes Are Derived from
                the Assumption of the Abundances of
                B Stars as the Interstellar Reference Abundances 
                and the Dust as Mixtures of 
                (i) Graphite + Fe-Containing Amorphous Silicates,
                (ii) Graphite + Fe-Lacking Amorphous Silicates 
                     + Iron Oxides, and
                (iii) Graphite + Fe-Lacking Amorphous Silicates 
                     + Iron.
    	        }
	\tablehead{
		\colhead{Star}&
		\colhead{$\AintobsWD/\NH$$^1$}&
		\colhead{$\AintobsWLJ/\NH$$^2$}&
		\multicolumn{3}{c}{Graphite\,+\,Fe-containing\,Silicate}&
		\colhead{}&
		\multicolumn{6}{c}{Graphite\,+\,Fe-lacking\,Silicate\,+\,Fe$_{x}$O$_{y}$}&
		\colhead{}&
		\multicolumn{4}{c}{Graphite\,+\,Fe-lacking\,Silicate\,+\,Fe}
		\\
		\cline{4-6}
		\cline{8-13}
		\cline{15-18}
		\\
		\colhead{}&
		\colhead{}&
		\colhead{}&
		\colhead{$\Vgra$/H$^3$}&
		\colhead{$\Vsil$/H$^4$}&
		\colhead{$\AintKK/\NH$$^5$}&
		\colhead{}&
		
		\colhead{$\Vgra$/H$^3$}&
		\colhead{$\Vsil$/H$^6$}&
		\colhead{$\VFeOa$/H$^7$}&
		\colhead{$\VFeOb$/H$^8$}&
		\colhead{$\VFeOc$/H$^9$}&
		\colhead{$\AintKK/\NH$$^5$}&
		\colhead{}&
		
		\colhead{$\Vgra$/H$^3$}&
		\colhead{$\Vsil$/H$^6$}&
		\colhead{$\VFe$/H$^{10}$}&		
		\colhead{$\AintKK/\NH$$^5$}
	}
\startdata
  HD 24912 & 1.24E-25 & 1.52E-25 & 1.51E-28 & 2.29E-27 & 7.41E-26 && 1.51E-28 & 1.73E-27 & 1.86E-28 & 2.24E-28 & 2.20E-28 & 7.46E-26 && 1.51E-28 & 1.74E-27 & 3.17E-28 & 6.46E-26\\
  HD 27778 & 9.07E-26 & 1.10E-25 & 1.15E-27 & 2.25E-27 & 1.22E-25 & & 1.15E-27 & 1.67E-27 & 1.91E-28 & 2.31E-28 & 2.26E-28 & 1.23E-25 & & 1.15E-27 & 1.67E-27 & 3.27E-28 & 1.13E-25\\
  HD 37021 & 1.42E-25 & 1.82E-25 & 1.05E-27 & 2.25E-27 & 1.17E-25 & &1.05E-27 & 1.68E-27 & 1.88E-28 & 2.27E-28 & 2.22E-28 & 1.18E-25 & & 1.05E-27 & 1.68E-27 & 3.21E-28 & 1.07E-25\\
  HD 37061 & 1.07E-25 & 1.35E-25 & 9.86E-28 & 2.38E-27 & 1.17E-25 && 9.86E-28 & 1.81E-27 & 1.92E-28 & 2.31E-28 & 2.27E-28 & 1.18E-25 && 9.86E-28 & 1.81E-27 & 3.27E-28 & 1.08E-25\\
  HD 116852 & 1.24E-25 & 1.50E-25 & 8.16E-28 & 2.19E-27 & 1.03E-25 & &8.16E-28 & 1.62E-27 & 1.85E-28 & 2.23E-28 & 2.19E-28 & 1.04E-25 && 8.16E-28 & 1.62E-27 & 3.16E-28 & 9.43E-26\\
  HD 122879 & 1.15E-25 & 1.42E-25 & -- & 2.12E-27 & 6.16E-26 & & -- & 1.53E-27 & 1.89E-28 & 2.27E-28 & 2.23E-28 & 6.28E-26 & & -- & 1.53E-27 & 3.21E-28 & 5.26E-26\\
  HD 147888 & 8.96E-26 & 1.13E-25 & 9.21E-28 & 2.28E-27 & 1.11E-25 & & 9.21E-28 & 1.70E-27 & 1.91E-28 & 2.30E-28 & 2.26E-28 & 1.12E-25 && 9.21E-28 & 1.70E-27 & 3.26E-28 & 1.02E-25\\
  HD 149757 & 1.47E-25 & 1.77E-25 & 9.62E-28 & 2.31E-27 & 1.14E-25 & & 9.62E-28 & 1.74E-27 & 1.90E-28 & 2.29E-28 & 2.25E-28 & 1.15E-25 & & 9.62E-28 & 1.74E-27 & 3.24E-28 & 1.05E-25\\
  HD 185418 & 1.16E-25 & 1.41E-25 & 3.68E-28 & 2.35E-27 & 8.62E-26 & & 3.68E-28 & 1.78E-27 & 1.90E-28 & 2.29E-28 & 2.24E-28 & 8.68E-26 && 3.68E-28 & 1.78E-27 & 3.24E-28 & 7.66E-26\\ 
  HD 207198 & 1.22E-25 & 1.48E-25 & 9.50E-28 & 2.26E-27 & 1.12E-25 & &9.50E-28 & 1.69E-27 & 1.89E-28 & 2.28E-28 & 2.24E-28 & 1.13E-25 && 9.50E-28 & 1.69E-27 & 3.23E-28 & 1.03E-25\\       	
	\enddata 
	\\
\begin{description}
\item[${1}$] $\AintobsWD/\NH$ (mag$\cm^3$\,H$^{-1}$)
     is the observed extinction (per H column)
     integrated from 912$\Angstrom$ to 1$\cm$, 
     with the extinction at $0.9 < \lambda < 1\cm$ 
     approximated by the $\RV=3.1$ model curve of 
     Weingartner \& Draine (2001);
\item[${2}$] $\AintobsWLJ/\NH$ (mag$\cm^3$\,H$^{-1}$)
     is the observed extinction (per H column)
     integrated from 912$\Angstrom$ to 1$\cm$, 
     with the extinction at $0.9 < \lambda < 1\cm$ 
     approximated by the $\RV=3.1$ model curve of 
     Wang, Li \& Jiang (2015a);
\item[${3}$] $\Vgra/\rmH$ (cm$^{3}$\,H$^{-1}$) is the volume 
     (per H nucleon) of graphite grains;
\item[${4}$] $\Vsil/\rmH$ (cm$^{3}$\,H$^{-1}$) is the volume 
     (per H nucleon) of Fe-containing amorphous silicate grains;
\item[${5}$] $\AintKK/\NH$ (mag$\cm^3$\,H$^{-1}$)
     is the KK-based wavelength-integrated extinction
     derived from the dust volumes;
\item[${6}$] $\Vsil/\rmH$ (cm$^{3}$\,H$^{-1}$) is the volume 
     (per H nucleon) of Fe-lacking amorphous silicate grains;
\item[${7}$] $\VFeOa/\rmH$ (cm$^{3}$\,H$^{-1}$) is the volume 
     (per H nucleon) of FeO grains;
\item[${8}$] $\VFeOb/\rmH$ (cm$^{3}$\,H$^{-1}$) is the volume 
     (per H nucleon) of Fe$_2$O$_3$ grains;
\item[${9}$] $\VFeOc/\rmH$ (cm$^{3}$\,H$^{-1}$) is the volume 
     (per H nucleon) of Fe$_3$O$_4$ grains; 
\item[${10}$] $\VFe/\rmH$ (cm$^{3}$\,H$^{-1}$) is the volume 
     (per H nucleon) of iron grains.
\end{description}
\end{deluxetable}

\thispagestyle{empty}
\setlength{\voffset}{25mm}
\begin{deluxetable}{lccccccccccccccccc}
	\rotate 
	\tablecolumns{18}
	\tabletypesize{\tiny}
	\tablewidth{0truein}
	\center
	\tablecaption{\footnotesize
		\label{tab:solar}
        Same as Table~\ref{tab:Bstar} 
        but with the Solar Abundances
        as the Interstellar Reference Abundances.
	}
	\tablehead{
		\colhead{Star}&
		\colhead{$\AintobsWD/\NH$}&
		\colhead{$\AintobsWLJ/\NH$}&
		\multicolumn{3}{c}{Graphite\,+\,Fe-containing\,Silicate}&
		\colhead{}&
		\multicolumn{6}{c}{Graphite\,+\,Fe-lacking\,Silicate\,+\,Fe$_{x}$O$_{y}$}&
		\colhead{}&
		\multicolumn{4}{c}{Graphite\,+\,Fe-lacking\,Silicate\,+\,Fe}
		\\
		\cline{4-6}
		\cline{8-13}
		\cline{15-18}
		\\
		\colhead{}&
		\colhead{}&
		\colhead{}&
		\colhead{$\Vgra$/H}&
		\colhead{$\Vsil$/H}&
		\colhead{$\AintKK/\NH$}&
		\colhead{}&
		
		\colhead{$\Vgra$/H}&
		\colhead{$\Vsil$/H}&
		\colhead{$\VFeOa$/H}&
		\colhead{$\VFeOb$/H}&
		\colhead{$\VFeOc$/H}&
		\colhead{$\AintKK/\NH$}&
		\colhead{}&
		
		\colhead{$\Vgra$/H}&
		\colhead{$\Vsil$/H}&
		\colhead{$\VFe$/H}&		
		\colhead{$\AintKK/\NH$}
	}
\startdata
  HD 24912 & 1.24E-25 & 1.52E-25 & 9.41E-28 & 2.47E-27 & 1.18E-25 & & 9.41E-28 & 1.81E-27 & 2.14E-28 & 2.58E-28 & 2.53E-28 & 1.19E-25 & & 6.85E-28 & 1.81E-27 & 3.66E-28 & 1.08E-25\\
  HD 27778 & 9.07E-26 & 1.10E-25 & 1.69E-27 & 2.43E-27 & 1.53E-25 & & 1.69E-27 & 1.74E-27 & 2.20E-28 & 2.65E-28 & 2.60E-28 & 1.55E-25& & 1.69E-27 & 1.74E-27 & 3.75E-28 & 1.43E-25\\
  HD 37021 & 1.42E-25 & 1.82E-25 & 1.58E-27 & 2.43E-27 & 1.48E-25 & & 1.58E-27 & 1.76E-27 & 2.17E-28 & 2.61E-28 & 2.56E-28 & 1.49E-25 & &1.58E-27 & 1.76E-27 & 3.70E-28 & 1.38E-25\\
  HD 37061 & 1.07E-25 & 1.35E-25 & 1.52E-27 & 2.56E-27 & 1.49E-25 & &1.52E-27 & 1.88E-27 & 2.20E-28 & 2.65E-28 & 2.61E-28 & 1.50E-25 & &1.52E-27 & 1.88E-27 & 3.76E-28 & 1.38E-25\\
  HD 116852 & 1.24E-25 & 1.50E-25 & 1.35E-27 & 2.37E-27 & 1.35E-25 & & 1.35E-27 & 1.70E-27 & 2.14E-28 & 2.58E-28 & 2.53E-28 & 1.36E-25 & & 1.35E-27 & 1.70E-27 & 3.65E-28 & 1.25E-25\\
  HD 122879 & 1.15E-25 & 1.42E-25 & -- & 2.30E-27 & 6.69E-26 & & -- & 1.61E-27 & 2.17E-28 & 2.62E-28 & 2.57E-28 & 6.86E-26 & & -- & 1.61E-27 & 3.70E-28 & 5.69E-26\\
  HD 147888 & 8.96E-26 & 1.13E-25 & 1.45E-27 & 2.46E-27 & 1.43E-25 & & 1.45E-27 & 1.78E-27 & 2.20E-28 & 2.65E-28 & 2.60E-28 & 1.44E-25 & &1.45E-27 & 1.78E-27 & 3.75E-28 & 1.32E-25\\            
  HD 149757 & 1.47E-25 & 1.77E-25 & 1.50E-27 & 2.49E-27 & 1.46E-25 & &1.50E-27 & 1.82E-27 & 2.19E-28 & 2.63E-28 & 2.58E-28 & 1.47E-25 & & 1.50E-27 & 1.82E-27 & 3.73E-28 & 1.35E-25\\
  HD 185418 & 1.16E-25 & 1.41E-25 & 9.01E-28 & 2.53E-27 & 1.18E-25 & & 9.01E-28 & 1.85E-27 & 2.19E-28 & 2.63E-28 & 2.58E-28 & 1.19E-25 & &9.01E-28 & 1.85E-27 & 3.73E-28 & 1.07E-25\\
  HD 207198 & 1.22E-25 & 1.48E-25 & 1.48E-27 & 2.44E-27 & 1.44E-25 & & 1.48E-27 & 1.77E-27 & 2.18E-28 & 2.62E-28 & 2.58E-28 & 1.45E-25 & &1.48E-27 & 1.77E-27 & 3.72E-28 & 1.33E-25\\
	
	\enddata 
	\\	
\end{deluxetable}

\thispagestyle{empty}
\setlength{\voffset}{25mm}
\begin{deluxetable}{lccccccccccccccccc}
	\rotate 
	\tablecolumns{18}
	\tabletypesize{\tiny}
	\tablewidth{0truein}
	\center
	\tablecaption{\footnotesize
		\label{tab:protosolar}
        Same as Table~\ref{tab:Bstar} 
        but with the Protosolar Abundances
        as the Interstellar Reference Abundances.
	}
	\tablehead{
		\colhead{Star}&
		\colhead{$\AintobsWD/\NH$}&
		\colhead{$\AintobsWLJ/\NH$}&
		\multicolumn{3}{c}{Graphite\,+\,Fe-containing\,Silicate}&
		\colhead{}&
		\multicolumn{6}{c}{Graphite\,+\,Fe-lacking\,Silicate\,+\,Fe$_{x}$O$_{y}$}&
		\colhead{}&
		\multicolumn{4}{c}{Graphite\,+\,Fe-lacking\,Silicate\,+\,Fe}
		\\
		\cline{4-6}
		\cline{8-13}
		\cline{15-18}
		\\
		\colhead{}&
		\colhead{}&
		\colhead{}&
		\colhead{$\Vgra$/H}&
		\colhead{$\Vsil$/H}&
		\colhead{$\AintKK/\NH$}&
		\colhead{}&
		
		\colhead{$\Vgra$/H}&
		\colhead{$\Vsil$/H}&
		\colhead{$\VFeOa$/H}&
		\colhead{$\VFeOb$/H}&
		\colhead{$\VFeOc$/H}&
		\colhead{$\AintKK/\NH$}&
		\colhead{}&
		
		\colhead{$\Vgra$/H}&
		\colhead{$\Vsil$/H}&
		\colhead{$\VFe$/H}&		
		\colhead{$\AintKK/\NH$}
	}
\startdata
    HD 24912 & 1.24E-25 & 1.52E-25 & 1.11E-27 & 2.91E-27 & 1.39E-25 & & 1.11E-27 & 2.20E-27 & 2.36E-28 & 2.84E-28 & 2.79E-28 & 1.40E-25 & & 1.11E-27 & 2.20E-27 & 4.03E-28 & 1.27E-25\\
    HD 27778 & 9.07E-26 & 1.10E-25 & 1.86E-27 & 2.87E-27 & 1.74E-25 & & 1.86E-27 & 2.13E-27 & 2.42E-28 & 2.91E-28 & 2.86E-28 & 1.75E-25 & & 1.86E-27 & 2.13E-27 & 4.12E-28 & 1.62E-25\\
   HD 37021 & 1.42E-25 & 1.82E-25 & 1.75E-27 & 2.86E-27 & 1.69E-25 & & 1.75E-27 & 2.14E-27 & 2.38E-28 & 2.87E-28 & 2.82E-28 & 1.70E-25 & &1.75E-27 & 2.14E-27 & 4.07E-28 & 1.57E-25\\
   HD 37061 & 1.07E-25 & 1.35E-25 & 1.69E-27 & 3.00E-27 & 1.70E-25 & & 1.69E-27 & 2.27E-27 & 2.42E-28 & 2.91E-28 & 2.86E-28 & 1.70E-25 & & 1.69E-27 & 2.27E-27 & 4.13E-28 & 1.57E-25\\
   HD 116852 & 1.24E-25 & 1.50E-25 & 1.52E-27 & 2.80E-27 & 1.56E-25 & & 1.52E-27 & 2.08E-27 & 2.36E-28 & 2.84E-28 & 2.79E-28 & 1.57E-25 & & 1.52E-27 & 2.08E-27 & 4.02E-28 & 1.44E-25\\  
   HD 122879 & 1.15E-25 & 1.42E-25 & -- & 2.73E-27 & 7.95E-26 & & -- & 2.00E-27 & 2.39E-28 & 2.88E-28 & 2.82E-28 & 8.08E-26 & & -- & 2.00E-27 & 4.07E-28 & 6.80E-26\\           
   HD 147888 & 8.96E-26 & 1.13E-25 & 1.62E-27 & 2.90E-27 & 1.64E-25 & & 1.62E-27 & 2.16E-27 & 2.42E-28 & 2.91E-28 & 2.85E-28 & 1.65E-25 & &1.62E-27 & 2.16E-27 & 4.12E-28 & 1.52E-25\\    
   HD 149757 & 1.47E-25 & 1.77E-25 & 1.66E-27 & 2.93E-27 & 1.67E-25 & &1.66E-27 & 2.20E-27 & 2.40E-28 & 2.89E-28 & 2.84E-28 & 1.67E-25 & & 1.66E-27 & 2.20E-27 & 4.10E-28 & 1.54E-25\\
   HD 185418 & 1.16E-25 & 1.41E-25 & 1.07E-27 & 2.96E-27 & 1.38E-25 & & 1.07E-27 & 2.24E-27 & 2.40E-28 & 2.89E-28 & 2.84E-28 & 1.39E-25 & & 1.07E-27 & 2.24E-27 & 4.10E-28 & 1.26E-25\\
   HD 207198 & 1.22E-25 & 1.48E-25 & 1.65E-27 & 2.88E-27 & 1.65E-25 & & 1.65E-27 & 2.15E-27 & 2.40E-28 & 2.88E-28 & 2.83E-28 & 1.65E-25 & & 1.65E-27 & 2.15E-27 & 4.08E-28 & 1.53E-25\\
	
	\enddata 
	\\	
\end{deluxetable}

\thispagestyle{empty}
\setlength{\voffset}{25mm}
\begin{deluxetable}{lccccccccccccccccc}
	\rotate 
	\tablecolumns{18}
	\tabletypesize{\tiny}
	\tablewidth{0truein}
	\center
	\tablecaption{\footnotesize
		\label{tab:protosolarGCE}
        Same as Table~\ref{tab:Bstar} 
        but with the GCE-augmented Protosolar Abundances
        as the Interstellar Reference Abundances.
	}
	\tablehead{
		\colhead{Star}&
		\colhead{$\AintobsWD/\NH$}&
		\colhead{$\AintobsWLJ/\NH$}&
		\multicolumn{3}{c}{Graphite\,+\,Fe-containing\,Silicate}&
		\colhead{}&
		\multicolumn{6}{c}{Graphite\,+\,Fe-lacking\,Silicate\,+\,Fe$_{x}$O$_{y}$}&
		\colhead{}&
		\multicolumn{4}{c}{Graphite\,+\,Fe-lacking\,Silicate\,+\,Fe}
		\\
		\cline{4-6}
		\cline{8-13}
		\cline{15-18}
		\\
		\colhead{}&
		\colhead{}&
		\colhead{}&
		\colhead{$\Vgra$/H}&
		\colhead{$\Vsil$/H}&
		\colhead{$\AintKK/\NH$}&
		\colhead{}&
		
		\colhead{$\Vgra$/H}&
		\colhead{$\Vsil$/H}&
		\colhead{$\VFeOa$/H}&
		\colhead{$\VFeOb$/H}&
		\colhead{$\VFeOc$/H}&
		\colhead{$\AintKK/\NH$}&
		\colhead{}&
		
		\colhead{$\Vgra$/H}&
		\colhead{$\Vsil$/H}&
		\colhead{$\VFe$/H}&		
		\colhead{$\AintKK/\NH$}
	}
\startdata
  HD 24912 & 1.24E-25 & 1.52E-25 & 1.56E-27 & 3.41E-27 & 1.76E-25 & & 1.56E-27 & 2.36E-27 & 3.28E-28 & 3.95E-28 & 3.88E-28 & 1.78E-25 & &1.56E-27 & 2.36E-27 & 5.60E-28 & 1.61E-25\\
  HD 27778 & 9.07E-26 & 1.10E-25 & 2.31E-27 & 3.37E-27 & 2.11E-25 & & 2.31E-27 & 2.29E-27 & 3.34E-28 & 4.02E-28 & 3.95E-28 & 2.14E-25 & & 2.31E-27 & 2.29E-27 & 5.70E-28 & 1.96E-25\\ 
  HD 37021 & 1.42E-25 & 1.82E-25 & 2.21E-27 & 3.36E-27 & 2.06E-25 & & 2.21E-27 & 2.30E-27 & 3.31E-28 & 3.98E-28 & 3.91E-28 & 2.09E-25 & & 2.21E-27 & 2.30E-27 & 5.64E-28 & 1.91E-25\\ 
  HD 37061 & 1.07E-25 & 1.35E-25 & 2.14E-27 & 3.50E-27 & 2.06E-25 & & 2.14E-27 & 2.43E-27 & 3.34E-28 & 4.03E-28 & 3.95E-28 & 2.09E-25 & & 2.14E-27 & 2.43E-27 & 5.70E-28 & 1.91E-25\\
  HD 116852 & 1.24E-25 & 1.50E-25 & 1.97E-27 & 3.30E-27 & 1.92E-25 & & 1.97E-27 & 2.25E-27 & 3.28E-28 & 3.95E-28 & 3.88E-28 & 1.96E-25 & & 1.97E-27 & 2.25E-27 & 5.59E-28 & 1.78E-25\\
  HD 122879 & 1.15E-25 & 1.42E-25 & 1.31E-28 & 3.23E-27 & 1.00E-25 & & 1.31E-28 & 2.16E-27 & 3.31E-28 & 3.99E-28 & 3.91E-28 & 1.04E-25 & & 1.31E-28 & 2.16E-27 & 5.65E-28 & 8.60E-26\\
  HD 147888 & 8.96E-26 & 1.13E-25 & 2.08E-27 & 3.40E-27 & 2.00E-25 & & 2.08E-27 & 2.33E-27 & 3.34E-28 & 4.02E-28 & 3.94E-28 & 2.03E-25 & & 2.08E-27 & 2.33E-27 & 5.69E-28 & 1.85E-25\\      
  HD 149757 & 1.47E-25 & 1.77E-25 & 2.12E-27 & 3.43E-27 & 2.03E-25 & & 2.12E-27 & 2.37E-27 & 3.33E-28 & 4.00E-28 & 3.93E-28 & 2.06E-25& & 2.12E-27 & 2.37E-27 & 5.67E-28 & 1.88E-25\\
  HD 185418 & 1.16E-25 & 1.41E-25 & 1.52E-27 & 3.46E-27 & 1.75E-25 & & 1.52E-27 & 2.40E-27 & 3.33E-28 & 4.00E-28 & 3.93E-28 & 1.78E-25 & &1.52E-27 & 2.40E-27 & 5.67E-28 & 1.60E-25\\
  HD 207198 & 1.22E-25 & 1.48E-25 & 2.11E-27 & 3.38E-27 & 2.01E-25 & & 2.11E-27 & 2.32E-27 & 3.32E-28 & 4.00E-28 & 3.92E-28 & 2.04E-25 & &2.11E-27 & 2.32E-27 & 5.66E-28 & 1.86E-25\\
	\enddata 
	\\
\end{deluxetable}

\thispagestyle{empty}
\setlength{\voffset}{25mm}
\begin{deluxetable}{lccccr}
\tablecolumns{5}
\tabletypesize{\scriptsize}
\tablewidth{0truein}
\center
\tablecaption{\footnotesize
		\label{tab:DustMaterials}
		Properties of Dust Materials}
\tablehead{
\colhead{Candidate}&
\colhead{Mass Density}&
\colhead{Static Dielectic Constant}&
\multicolumn{2}{c}{$F(a/b;\varepsilon_{0}$)}
\\
\cline{4-5}
\colhead{Dust Material}&
\colhead{$\rho$ (g$\cm^{-3}$)}&
\colhead{($\varepsilon_{0}$)}&
\colhead{Prolate ($a/b=3$)}&
\colhead{Oblate ($a/b=1/2$)}
}

\startdata	
Forsterite (Mg$_{2}$SiO$_{4}$) & 3.27$^{a}$ &5.5$^{a}$ & 0.668 & 0.633\\
Enstatite (MgSiO$_{3}$) & 3.2$^{b}$ &6.7$^{b}$ & 0.749 & 0.698\\
Olivine (MgFeSiO$_{4}$) & 3.5$^{a}$ &10$^{a}$ & 0.905 & 0.814\\
W\"ustite (FeO) & 5.7$^{c}$ &24$^{c}$ & 1.18 & 0.989\\
Haematite ($\alpha$-Fe$_{2}$O$_{3}$) & 5.26$^{d}$&16$^{e}$&1.07&0.920\\
Magnetite (Fe$_{3}$O$_{4}$)&5.18$^{d}$&$\infty$&1.52&1.15\\
Iron (Fe)&7.8&$\infty$&1.52&1.15\\
Graphite (C)&2.24&$\infty$&1.52&1.15\\
\enddata 
\\
(a) J\"ager et al.\ (2003);
(b) Dorschner et al.\ (1995);
(c) Henning \& Mutschke (1997);
(d) Schrettle et al.\ (2012);
(e) Steyer (1974).
\end{deluxetable}

\end{document}